\PassOptionsToPackage{svgnames}{xcolor}
\documentclass[sigconf,nonacm,natbib=false,screen]{acmart}

\usepackage[utf8]{inputenc}
\usepackage[T1]{fontenc}
\usepackage[american]{babel}

\usepackage[frozencache]{minted}
\usepackage{csquotes} % Should be loaded after minted/fancyvrb

\usepackage{booktabs}
\usepackage{listings}
\usepackage{enumitem}
\usepackage{amsmath}
\usepackage{subcaption}
\usepackage{newfloat}
\usepackage{tabularx}
\usepackage{makecell}
\usepackage{environ}
\usepackage[disable]{todonotes}   % notes showed
\usepackage{xspace}
\usepackage{ifthen}
\usepackage[normalem]{ulem}
\usepackage{lipsum}
\usepackage{afterpage}

\DeclareFloatingEnvironment[
  name={Listing},
]{flisting}

\DeclareFloatingEnvironment[
  fileext=lop,
  listname={List of Programs},
  name=Program
]{program}
\DeclareCaptionSubType{program}

\usepackage[binary-units]{siunitx}
\DeclareSIUnit{\nothing}{\relax}
\DeclareSIUnit{\x}{\times}
\usepackage{tikz}
\usetikzlibrary{fit}
\usetikzlibrary{positioning}

\useshorthands*{"}
\defineshorthand{"-}{\babelhyphen{hard}}

\PassOptionsToPackage{
  sortcites,
  maxcitenames=2,
  date=year,
  abbreviate=true,
  backend=biber,
  backref=false,
  dateabbrev=true,
  doi=true,
  isbn=true,
  language=american,
  maxbibnames=9,
  style=ACM-Reference-Format,
  url=true,
  urldate=comp,
}{biblatex}
\usepackage{biblatex}
\bibliography{main}

\definecolor{ETHa}{RGB}{31,64,122}      % ETH1 (dark blue)
\definecolor{ETHb}{RGB}{72,90,44}       % ETH2 (dark green)
\definecolor{ETHc}{RGB}{18,105,176}     % ETH3 (light blue)
\definecolor{ETHd}{RGB}{114,121,28}     % ETH4 (light green)
\definecolor{ETHe}{RGB}{145,5,106}      % ETH5 (dark pink)
\definecolor{ETHf}{RGB}{111,111,100}    % ETH6 (gray)
\definecolor{ETHg}{RGB}{168,50,45}      % ETH7 (dark red)
\definecolor{ETHh}{RGB}{0,122,150}      % ETH8 (turquoise)
\definecolor{ETHi}{RGB}{149,96,19}      % ETH9 (brown)

\PassOptionsToPackage{pdfpagelabels=false}{hyperref}
\usepackage{hyperref}

\hypersetup{%
	colorlinks=true,%
	urlcolor=ETHg,%
	linkcolor=ETHc,%
	citecolor=ETHd,%
	breaklinks=true,%
	pageanchor=true,%
}

\makeatletter
\DeclareRobustCommand{\varname}[1]{\begingroup\newmcodes@\mathit{#1}\endgroup}
\makeatother

\newcommand\mdots{\hbox to.5em{\hss.\hss\hss.\hss\hss.\hss}}

\newcommand{\queryref}[1]{(\hyperref[benchmark:q#1]{Q#1})\xspace}

\newcounter{requirementone}
\newcounter{requirementtwo}
\newcounter{requirementthree}
\newcommand{\requirement}[2]{%
  \ifnum #1=1%
    \mbox{\refstepcounter{requirementone}(R#1.\therequirementone\label{r#1:#2})}\xspace%
  \fi%
  \ifnum #1=2%
    \mbox{\refstepcounter{requirementtwo}(R#1.\therequirementtwo\label{r#1:#2})}\xspace%
  \fi%
  \ifnum #1=3%
    \mbox{\refstepcounter{requirementthree}(R#1.\therequirementthree\label{r#1:#2})}\xspace%
  \fi%
}
\newcommand{\requirementref}[2]{(\hyperref[r#1:#2]{R#1.\ref{r#1:#2}})\xspace}

\newcommand\vldbdoi{10.14778/3489496.3489498}
\newcommand\vldbpages{154 - 168}
\newcommand\vldbvolume{15}
\newcommand\vldbissue{2}
\newcommand\vldbyear{2022}
\newcommand\vldbauthors{\authors}
\newcommand\vldbtitle{\shorttitle}

\newcommand\myshortauthors{D. Graur, I. Müller, M. Proffitt,
                           G. Fourny, G. T. Watts, G. Alonso}

\begin{document}

\title[Evaluating Query Languages and Systems for High-Energy Physics Data]
      {Evaluating Query Languages and Systems \\ for High-Energy Physics Data}
\subtitle{[Extended Version]}

\author{Dan Graur}
\affiliation{%
  \institution{Department of Computer Science}
  \institution{ETH Zurich}
  \streetaddress{Stampenbachstrasse 114}
%  \city{Zurich}
%  \state{Switzerland}
  \postcode{8057}
}
\email{dan.graur@inf.ethz.ch}

\author{Ingo Müller}
\orcid{0000-0001-8818-8324}
\affiliation{%
  \institution{Department of Computer Science}
  \institution{ETH Zurich}
  \streetaddress{Stampenbachstrasse 114}
%  \city{Zurich}
%  \state{Switzerland}
  \postcode{8057}
}
\email{ingo.mueller@inf.ethz.ch}

\author{Mason Proffitt}
\orcid{0000-0001-8740-8866}
\affiliation{%
  \institution{Department of Physics}
  \institution{University of Washington}
%  \city{Seattle}
%  \state{United States}
}
\email{masonLp@uw.edu}

\author{Ghislain Fourny}
\orcid{0000-0001-8740-8866}
\affiliation{%
  \institution{Department of Computer Science}
  \institution{ETH Zurich}
  \streetaddress{Stampenbachstrasse 114}
%  \city{Zurich}
%  \state{Switzerland}
  \postcode{8057}
}
\email{ghislain.fourny@inf.ethz.ch}

\author{Gordon T. Watts}
\affiliation{%
  \institution{Department of Physics}
  \institution{University of Washington}
%  \city{Seattle}
%  \state{United States}
}
\email{gwatts@uw.edu}

\author{Gustavo Alonso}
\affiliation{%
  \institution{Department of Computer Science}
  \institution{ETH Zurich}
  \streetaddress{Stampenbachstrasse 114}
%  \city{Zurich}
%  \state{Switzerland}
  \postcode{8057}
}
\email{alonso@inf.ethz.ch}

\begin{abstract}
In the domain of high-energy physics (HEP), query languages in general and SQL in particular have found limited acceptance.
This is surprising since HEP data analysis matches the SQL model well:
the data is fully structured and queried using mostly standard operators.
To gain insights on why this is the case, we perform a comprehensive analysis
of six diverse, general-purpose data processing platforms
using an HEP benchmark.
The result of the evaluation is an interesting
and rather complex picture of existing solutions:
Their query languages vary greatly in how natural and concise
 HEP query patterns can be expressed.
Furthermore, most of them are also
between one and two orders of magnitude slower
than the domain-specific system used by particle physicists today.
These observations suggest that,
while database systems and their query languages
are \emph{in principle} viable tools for HEP,
significant work remains to make them relevant to HEP researchers.

\end{abstract}

\maketitle

\medskip
\begingroup\small\noindent\raggedright\textbf{Full Paper Reference:}\\
This document is the extended version of the following full paper:\\
\vldbauthors. \vldbtitle. PVLDB, \vldbvolume(\vldbissue): \vldbpages, \vldbyear.\\
\href{https://doi.org/\vldbdoi}{doi:\vldbdoi}
\endgroup

\makeatletter
\fancyfoot[C]{\footnotesize\thepage}%
\fancyhead[L]{\ACM@linecountL\@headfootfont\shorttitle}%
\fancyhead[R]{\@headfootfont\myshortauthors\ACM@linecountR}%
\makeatother

\begingroup
\renewcommand\thefootnote{}\footnote{\noindent
  This work is licensed under the Creative Commons
  Attribution-NonCommercial-NoDerivatives (BY-NC-ND) 4.0
  International License.
  To view a copy of this license,
  visit \url{http://creativecommons.org/licenses/by-nc-nd/4.0/}.
  For any use beyond those covered by this license,
  obtain permission by contacting the authors.
  Copyright is held by the owner/author(s).
}\addtocounter{footnote}{-1}
\endgroup

\section{Introduction}
\label{sec:introduction}

In the domain of High-Energy Physics (HEP),
the well-known advantages of data processing platforms (data independence, declarative language, etc.) have not led to their adoption.
This is surprising given the nature of the data and queries
typical for that domain:
Data sets in HEP are large but always fully structured.
However, they are also heavily nested:
they represent ``events'' registered
by the sensors of a particle collider,
where each event consists of a few scalar attributes
as well as of numerous variable-sized sequences of relatively wide records.
Queries follow a relatively simple pattern:
they typically consist of a single scan over the input
involving only a small subset of the available attributes,
derivation of additional measures
(potentially by joining and reducing
the sequences \emph{within the same event}),
and selection of an interesting subset of events,
which are then summarized using a reduction.
HEP data is thus stored and analyzed
in non-first normal form (NF$^\text{2}$)---%
a feature that early database systems did not support
and thus the main reason why relational engines
were rejected by physicists historically %
(along with the lack of support for used-defined code~\cite{Malon:1997hq}).

Nowadays, most particle physicists work with a domain-specific system
called the ROOT framework~\cite{BRUN199781,Antcheva2009},
and increasingly so with its new RDataFrame interface~\cite{rdataframes}.
In ROOT, queries are written in C++, requiring a non-trivial user effort,
which can deter less experienced users~\cite{Gutsche_2017}.
Queries in ROOT entangle many aspects of the storage format and file system,
the in-memory runtime format, the execution strategy,
the target platform, and even the visualization.
This makes many of the proven techniques known from data management
difficult or impossible to apply.
As one consequence, since ROOT does not support
\emph{distributed} processing out of the box,
there is no standard scale-out solution for HEP analyses
and various groups of physicists have built their own solutions
for splitting up jobs into tasks, scheduling them across clusters,
and combining the results.
At the same time, many relational systems today
offer rather complete support for nested data types
like variable-size arrays, suggesting that the question
of whether these general-purpose data processing systems
are suitable or not for HEP should be revisited.

In this paper, we perform a comprehensive analysis
in terms of expressiveness and performance
of six general-purpose data processing systems
that are \emph{in principle} suited for HEP analyses:
Postgres~\cite{StonebrakerMichael1986},
as a representative
of conventional database systems,
Presto~\cite{sethi2019presto},
a representative of systems for distributed data analytics
with support for nested data,
Google BigQuery~\cite{sato2012inside}
and Amazon Athena~\cite{documentation2018amazon}
as two Query-as-a-Service systems designed for large-scale analytics,
as well as AsterixDB~\cite{10.14778/2733085.2733096}
and RumbleDB~\cite{10.14778/3436905.3436910},
two scale-out systems designed for document-oriented analytics.
The baseline consists of the RDataFrames interface.
We use the Analysis Description Languages (ADL)
benchmark~\cite{hepadlbenchmarks},
created by physicists to evaluate languages and systems in their domain,
and analyze typical query patterns occurring in the queries of the benchmark.
We identify 16 language features that are useful
in implementing these patterns,
categorize them in terms of how essential they are,
and analyze how well the different query languages implement them.

The result is a complex and rather intriguing picture:
The two document-oriented systems, AsterixDB and RumbleDB, allow for the most natural and succinct query implementations.
Their languages often even seem more elegant
than those using RDataFrames,
and, more importantly, their declarative nature also avoids
most of the drawbacks of the ROOT framework mentioned above.
They thus have the potential
to increase the productivity of physicists significantly.
To a slightly lesser degree, the same is true
for the SQL dialects of BigQuery and Postgres,
which implement the relevant parts of the SQL standard
related to composite types and arrays.
The conclusion is, hence, that there is no reason anymore
to exclude SQL as a language for HEP analyses \emph{per se}---%
there are, however, significant weaknesses in some of its dialects:
Those of Athena and Presto are the most limited in the study;
however, unlike many other systems that we excluded from the analysis,
they \emph{are} expressive enough to implement the queries even if it requires some effort.
In terms of performance, the ranking is roughly reversed:
AsterixDB, RumbleDB, and Postgres
are the slowest systems in the comparison,
about an order of magnitude slower
than the fastest general-purpose system, Presto,
which is another order of magnitude slower than RDataFrames.
While their languages are well suited,
these systems are hence not efficient enough to be practical.
BigQuery has generally the lowest running times
but the cost of individual queries is often an order of magnitude higher
than that of RDataFrames.
Overall, our study reveals significant weaknesses
in all analyzed data processing systems in
what could be considered their core competence
and explains why physicists have resorted to developing their own solution.

% The following section should contain 2 subsections: 'the data' and 'the queries'
\section{High-energy Physics Data Analysis}
\label{sec:the-dataset}
\subsection{High-energy Physics}

High-energy physics (or particle physics)
studies the nature of the particles that constitute matter and radiation by observing the collisions of such particles in accelerators
such as the Large Hadron Collider (LHC) run by CERN. 

The nature of the particles
and the need for high statistical confidence
imply that a large number of similar collisions have to be analyzed.
This makes the HEP community no stranger to large-scale data analytics.
For example, the LHC in its latest configuration (Run 2),
makes two particle beams intersect at a rate of \SI{40}{\mega\hertz},
thus producing 40 million so-called ``events'' per second~\cite{WikipediaLHC}.
In each event, sensors
register the presence and paths of particles resulting from the collision.
The events are filtered on-site by a cascade of automatic filters
until eventually about 1000 events per second are archived and shared%
~\cite{WikipediaCMS}
and subsequently analyzed by particle physicists around the world---%
the process we focus on in this paper.
The amount of data archived is staggering:
Run 2 produced data at a rate of about \SI{8}{\giga\byte\per\second},
which amounted to \SI{88}{\peta\byte} in 2018,
and the next configurations,
Run 3 and Run 4, scheduled for 2021 and 2027,
will produce \SI{2}{\times} and \SI{10}{\times} that amount%
~\cite{LhcAbout,Meglio2017,Calafiura2020}.

\begin{flisting}
\begin{minted}{c++}
struct MET  { float pt, phi, sumet /*...*/; };
struct Muon { float pt, eta, phi, mass /*...*/;
              int charge; };
struct Electron { /*...*/ };
// ...
struct Event { int event, run; MET met; // ...
    vector<Muon> muons;
    vector<Electron> electrons;
    /*...*/ };
\end{minted}
  \vspace{-2ex}
  \caption{Simplified schema of typical HEP data sets.}
  \label{lst:hep-schema}
\end{flisting}

The information describing each event
is shown in Listing~\ref{lst:hep-schema}.
It consists of per-event metadata
such as the run ID and the event ID,
per-event measurements such as the missing energy (\mintinline{c++}{MET}),
and information about various types of observed particles.%
\footnote{For simplicity  we denote particle-like objects such as jets
          simply as ``particles.''}
Which particle types may occur
depends on the underlying collision experiment
as well as prior processing of the data set (called ``reconstruction'')
and include jets, electrons, muons, tau particles, and photons.
All particle types have a common set of dimensions
including transverse momentum (\mintinline{c++}{pt}),
pseudorapidity (\mintinline{c++}{eta}),
azimuth angle (\mintinline{c++}{phi}), and mass,
but some particle types have additional dimensions.
For instance, electrons have a charge while jets do not.
In each event, zero, one, or more particles
may be observed of each type.

High-energy physics data is thus fully structured, contains no \mintinline{SQL}{NULL} values,
and could be stored in normalized form in any RDBMS
using one table per particle type with foreign keys to an event table.
However, there is a strong ownership relationship
between an event and its particles,
and the particles are not analyzed
outside of the context of their event.
Physicists thus store HEP data \emph{always}
in non-first normal form (NF$^2$),
i.e., particles are stored nested as part of each event.
This makes it possible to store events in files
without ever breaking foreign key constraints
and eliminates both the mental effort and the execution cost of joins.

Data sets usually contain a large number of attributes
(all available dimensions of all potentially present particle types,
both measured or derived),
at least several dozen and sometimes in the thousands.
However, each query typically only accesses a few of them
depending on what aspect the physicist is currently interested in.
The data formats used for HEP data are thus all columnar formats
in order to allow for pushing projections into the storage layer.
For better or for worse, this columnar representation
is also exposed by most programming abstractions
commonly used for HEP analyses,
which we discuss in more detail below.

In this paper, we concentrate on the final processing step (``analysis'').
Virtually all queries in this step consist of two phases:
(1) a sequence of transformations and filters
applied \emph{to each event in isolation},
and (2) one or several aggregations of the remaining events.
The first phase consists of tasks such as
establishing the presence of a particle type
that sensors cannot detect directly
but that can be derived from the presence of other particles
or selecting events that contain
a particular combination of interesting particles.
The second phase consists of summarizing the selected events
in form of a histogram of one or several dimensions of these events.

The analysis starts by determining some basic properties of the data set
to confirm whether the data set
is a good candidate for further exploration.
Then, increasingly complex patterns are tried out,
each of which is plotted in increasing detail
 to steer the subsequent search,
until the result (i.e., the query itself as well as the plots)
is eventually shared as code or as a materialized view.

\subsection{The ADL Benchmark}

The \emph{Institute for Research and Innovation in Software
for High Energy Physics} (IRIS-HEP)~\cite{IrisHepWebsite}
has recently published the Analysis Description Languages (ADL)
benchmark~\cite{hepadlbenchmarks},
a sequence of high-level query descriptions
designed to represent typical patterns in HEP data analysis.
Its goal is to facilitate the test and comparison of languages and systems,
and thus to guide the design of next-generation tools in the HEP domain.
We use this benchmark in v0.1 as a running example
for the remainder of the paper.

\label{sec:the-dataset:data}
The data set of the benchmark consists of data obtained
as a result of the Compact Muon Solenoid (CMS) experiment
run on the Large Hadron Collider in the year 2012%
~\cite{CmsOpenData2012}.
It consists of roughly 54 million events (i.e., rows)
and a total of 65 attributes (i.e., columns),
totaling approximately \SI{17}{\giga\byte} in the ROOT format.

\label{sec:the-dataset:queries}
The benchmark consists of the following eight queries.
We make slight modifications to the query text
in order to disambiguate some physics terminology:
\begin{enumerate}[label=(Q\arabic*)]
    \item\label{benchmark:q1}
      Plot the $E_T^{miss}$ (missing transverse energy) of all events.
    \item\label{benchmark:q2}
      Plot the $p_T$ (transverse momentum) of all jets in all events.
    \item\label{benchmark:q3}
      Plot the $p_T$ of jets with $|\eta| < 1$ (jet pseudorapidity).
    \item\label{benchmark:q4}
      Plot the $E_T^{miss}$ of the events
      that have at least two jets
      with $p_T > \SI{40}{\giga\electronvolt}$ (gigaelectronvolt).
    \item\label{benchmark:q5}
      Plot the $E_T^{miss}$ of events
      that have an opposite-charge muon pair
      with an invariant mass between
      \SI{60}{\giga\electronvolt} and \SI{120}{\giga\electronvolt}.
    \item\label{benchmark:q6}\label{benchmark:q6a}\label{benchmark:q6b}
      For events with at least three jets,
      plot the $p_T$ of the trijet system four-momentum
      (i.e., any combination of three distinct jets within the same event)
      that has the invariant mass closest to \SI{172.5}{\giga\electronvolt}
      in each event
      and plot the maximum b-tagging discriminant value
      among the jets in this trijet.
    \item\label{benchmark:q7}
      Plot the scalar sum in each event of the $p_T$ of the jets
      with $p_T > \SI{30}{\giga\electronvolt}$
      that are not within 0.4 in $\Delta R$ of any light lepton
      (i.e., electron or muon) with $p_T > \SI{10}{\giga\electronvolt}$.
    \item\label{benchmark:q8}
      For events with at least three light leptons
      and a same-flavor opposite-charge light lepton pair,
      find such a pair that has the invariant mass
      closest to \SI{91.2}{\giga\electronvolt} in each event
      and plot the transverse mass of the system,
      consisting of the missing transverse momentum
      and the highest-$p_T$ light lepton not in this pair.
\end{enumerate}

Note that ``to plot'' is short for
``to plot an appropriate equi-width histogram,''
where 100 is a typical number of bins,
the highest and lowest bins are typically set statically
based on domain knowledge about the plotted metric,
and under- and over-flows have their dedicated bins.
Also, note that \queryref{6} consists of two different plots,
i.e., a common sequence of transformations and filters
is consumed by two distinct aggregations, as discussed above.
We refer to the two plots as \queryref{6a} and \queryref{6b}, respectively.
Finally, note that some properties such as the mass of a trijet
are calculated using involved mathematical formulae,
which we do not include for conciseness.
We refer to our query implementations~\cite{BenchmarkScripts} for details.

% The following section should contain the a discussion of the implementations across the chosen systems, as well as grievances and positive points. 
\section{Analysis of HEP Query Patterns}
\label{sec:implementation}
We start with an analysis of functional requirements
for data analytics in HEP
with a focus on the query language.
We interleave this analysis with a survey
of existing general-purpose data processing systems
that are, at least in principle, suitable for the HEP domain.

\subsection{Methodology}

\textbf{Classification of requirements}. 
We define the following classes of functional requirements:
\begin{enumerate}[label=(R\arabic*)]
  \item Essential functionality,
    without which HEP queries cannot be reasonably expressed.
  \item Important functionality,
    which has a major impact on readability and conciseness
    but acceptable alternatives exist.
  \item Useful functionality helping to improve code quality
    but with limited impact.
\end{enumerate}
In the analysis below, we label each requirement with (R$i$.$j$),
where $i$ denotes one of the classes and $j$ identifies the requirement.
It is difficult to make such a classification fully objective
as it depends on the user's taste, past experience, etc.
However, it does help to interpret and summarize the findings.

\textbf{Systems}.
We implement the queries in several SQL dialects
as well as the documnet-oriented languages JSONiq and SQL++,
and contrast them with state-of-the-art approach in the HEP domain,
the\emph{RDataFrames} interface of the ROOT framework,
of which we use v6.24.02~\cite{rene_brun_2019_3895860}.
For SQL, we consider the dialects of
\emph{PrestoDB}~\cite{sethi2019presto} v0.258
(or \emph{Presto} for short),
a modern system for large-scale analytics,
\emph{PostgreSQL}~\cite{StonebrakerMichael1986} version 13
(or \emph{Postgres} for short),
as a full-featured representative
of conventional database systems,
\emph{Amazon Athena}~\cite{documentation2018amazon} (engine version 2),
a Query-as-a-Service system based on PrestoDB,
and \emph{Google BigQuery}~\cite{sato2012inside}, a Query-as-a-Service system
built as the public version of Dremel~\cite{melnik2010dremel}.
We use PrestoDB rather than its fork \emph{Trino}
(previously called Presto SQL)
because the latter does not support SQL-based user-defined functions.
Presto does not make any claims of complying with the SQL standard,
Postgres implements most features of SQL:2003
but does not aim at exact conformance,
Athena uses ``standard SQL'' according to the product description,
and BigQuery uses SQL:2011 with a few extensions.

We use JSONiq and SQL++ because they are designed
to deal with the nested (and heterogeneous) JSON data model,
which captures all aspects of HEP data naturally.
We use our own implementation of JSONiq,
\emph{RumbleDB}~\cite{10.14778/3436905.3436910} v1.11.0,
which is built atop Apache Spark;
however, the JSONiq queries run without modification
on the two independent JSONiq implementations
Zorba~\cite{zorba} and Xidel~\cite{xidel} as well.
Both these systems are single-threaded
and hence not optimal for HEP applications.
SQL++ is the main query language of AsterixDB~\cite{10.14778/2733085.2733096},
a large-scale document store.
We use the development version of AsterixDB with git hash \texttt{81c32493},
which is about two months older than the released version 0.9.7
and includes the fix to a performance bug we found during this study.

We also considered a number of additional systems;
however, many of them lack the support of even the most basic features,
so we excluded them from the comparison.
For example, MySQL~\cite{MySQL} and hence its hosted versions
Amazon Aurora~\cite{10.1145/3035918.3056101}
and MySQL HeatWave~\cite{MySqlHeatWave},
MS SQL Server~\cite{SqlServer} and hence
Synapse Analytics~\cite{SynapseAnalytics} (which uses the same SQL dialect),
and Actian Vector~\cite{ActianVector}
do not have support for arrays
and can thus not even represent the input data
(unless normalized, which we exclude for the reasons given above).
SAP HANA~\cite{Farber2012},
and via their JSON type also MonetDB~\cite{Boncz2008}
and Snowflake~\cite{10.1145/2882903.2903741},
do have an array type but no suitable construct
for unnesting or otherwise querying their elements.
We also considered Spark SQL~\cite{Armbrust2015}
but, in the end, decided to leave it as future work:
it does support arrays and structs,
but can only query the former through a set of array functions,
against which we present arguments below.
In contrast, we believe that the SQL dialect
of \emph{Apache Drill}~\cite{Hausenblas2013}
as well as the query language \emph{PartiQL}~\cite{Partiql2019}
(based on SQL++, designed by AWS,
and available for \emph{Amazon Redshift}~\cite{Redshift})
are suitable candidates for HEP analysis.
Due to time and space constraints,
we leave them  as future work as well.

\todo[inline]{Other candidates are:
  - Apache Drill (which is designed for in-situ analysis of heterogeneous data),
  - Oracle Database (which is probably the most complete DBMS out there). \\
  - Amazon Aurora (Postgres flavor)
    * I *think* that this has array support, since the documentation mentions the Postgres parameter "array\_nulls"
    https://docs.aws.amazon.com/AmazonRDS/latest/AuroraUserGuide/AuroraPostgreSQL.Reference.html
  Excluded because array support is missing:
  - MySQL
  - Actian Vector: https://docs.actian.com/vector/6.0/index.html#page/Connectivity/Data_Type_Compatibility.htm
  - MS SQL server (and hence Synapse Analytics, which also uses T-SQL)
    * no array support: https://stackoverflow.com/a/41343387/651937
  - Redshift
    * the documentation says that arrays are converted to varchar(64k)
    * It can query Parquet from S3 and has another query language, PartiQL,
      which is based on SQL++ and should thus work similarly well.
    * https://aws.amazon.com/blogs/opensource/announcing-partiql-one-query-language-for-all-your-data/
https://docs.aws.amazon.com/redshift/latest/dg/federated-data-types.html
  Other excluded systems:
  - Snowflake:
    * has support for arrays
      https://docs.snowflake.com/en/sql-reference/data-types-semistructured.html
    * does not have support subqueries
    * supports unnesting of arrays via lateral join
    * supports array functions, but misses "map", "filter", "join", etc
      https://docs.snowflake.com/en/sql-reference/functions-semistructured.html
  - SAP HANA:
    * Has arrays but no sub-queries
    * https://help.sap.com/viewer/de2486ee947e43e684d39702027f8a94/2.0.02/en-US/cba8ef91ba944e37beb26eb8bd995c2f.html
  - MonetDB:
    * Does not seem to supports arrays: https://www.monetdb.org/Documentation/SQLLanguage/DataTypes/BuiltinSQLTypes
    * Documentation mentions "array size": https://www.monetdb.org/wiki/MonetDB\_type\_system
    * Support JSON type: https://www.monetdb.org/Documentation/SQLLanguage/DataTypes/JSONDataType
    * Does not support UNNEST: https://www.monetdb.org/Documentation/SQLLanguage/Queries/TableExpressions}

\textbf{Data Format}.
Since none of the general-purpose systems
supports reading from ROOT files directly,
we convert the data to Parquet,
which can represent ROOT files accurately
and achieves a similar compression ratio.
In this process, we also convert the data to a more natural representation:
While the original ROOT files
decompose the fields of structured attributes
into distinct columns, both physically \emph{and} logically,
we represent them as logical \texttt{struct}s and arrays thereof,
which are both physically decomposed into scalar columns by Parquet.
Instead of having to re-compose particle arrays from various attributes
such as \texttt{Jet\_pt}, \texttt{Jet\_eta}, etc. and \texttt{nJet},
queries thus simply access the \texttt{Jets} attribute
of type \texttt{array<struct<float pt, float eta, \mdots>{}>}.
Thanks to the separation of logical and physical representation of the data,
the data is exposed in a more intuitive way to the user
while maintaining the performance benefits of column decomposition.

\subsection{Accessing and Creating Nested Structs}

\begin{flisting}
  \centering
  \begin{subflisting}{\linewidth}
    \begin{minted}{c++}
df.Define("Jet_p4", make_p4, {"Jet_pt", "Jet_eta",
                              "Jet_phi", "Jet_mass"})
    \end{minted}
    \vspace{-2ex}
    \caption{RDataFrames.}
    \vspace{1.5ex}
    \label{alg:struct:rdf}
  \end{subflisting}
  \begin{subflisting}{\linewidth}
    \begin{minted}{SQL}
STRUCT<x INT64, y FLOAT64>(a.x + b.x, 42.0)),
STRUCT(a.x + b.x AS x, 42.0)
    \end{minted}
    \vspace{-1ex}
    \caption{BigQuery.}
    \vspace{1.5ex}
    \label{alg:struct:bigquery}
  \end{subflisting}
  \begin{subflisting}{\linewidth}
    \begin{minted}{SQL}
CAST(ROW(a.x + b.x, 42.0) AS ROW(x BIGINT, y DOUBLE))
    \end{minted}
    \vspace{-1ex}
    \caption{Presto.}
    \vspace{1.5ex}
    \label{alg:struct:presto}
  \end{subflisting}
  \begin{subflisting}{\linewidth}
    \begin{minted}{SQL}
CAST(ROW((a).x + (b).x, 42.0) AS userDefinedPair)
    \end{minted}
    \vspace{-1ex}
    \caption{Postgres.}
    \vspace{1.5ex}
    \label{alg:struct:postgres}
  \end{subflisting}
  \begin{subflisting}{\linewidth}
    \begin{minted}{xquery}
{ "x": $a.x + $b.x, "y": 42.0 }
    \end{minted}
    \vspace{-1ex}
    \subcaption{JSONiq.}
    \label{alg:struct:jsoniq}
    \vspace{1.5ex}
  \end{subflisting}
  \begin{subflisting}{\linewidth}
    \begin{minted}{SQL}
SELECT {"x": a.x + b.x, "y": 42.0};
SELECT (SELECT a.x + b.x AS x, 42.0 AS y);
    \end{minted}
    \vspace{-1ex}
    \caption{SQL++.}
    \label{alg:struct:sqlpp}
  \end{subflisting}
  \vspace{-3ex}
  \caption{Accessing and creating nested structs.}
  \label{alg:struct}
\end{flisting}

We now analyze basic language constructs
for manipulating nested structs (or ``objects''),
starting with the availability
of \dotuline{\requirement{2}{structs} structured data types}.
This is a useful feature for handling several related attributes jointly,
such as the various dimensions describing individual particles,
and is thus ubiquitous in HEP queries.
RDataFrames have no explicit support for structs but,
since they allow arbitrary C++ types,
users can create structured types manually
or use existing types as they see fit.
A typical example consists of assembling the library type
for 4-dimensional space-time vectors
(called ``Lorentz-vectors'')
from their constituent atomic values (Listing~\ref{alg:struct:rdf}).
The function \mintinline{c++}{make_p4}
is written by the user in C++ for the purpose of this query
and applies the construction of Lorentz-vectors to arrays.
Input data from ROOT files, however, is exposed in a columnar data model
and not in the form of structs.
Queries thus typically contain many columnar operations,
where the correspondence between the dimensions of individual particles
is given only implicitly by matching indices of different columns.
How this looks like becomes clearer with the examples below.

In contrast, the SQL standard describes
the \mintinline{SQL}{ROW} type, as well as user-defined (composite) types,
which were introduced in SQL:1999 for this purpose.
BigQuery implements the \mintinline{SQL}{ROW} type
under the type name \mintinline{SQL}{STRUCT}.
As the first expression in Listing~\ref{alg:struct:bigquery} shows,
an instance can be constructed
by \dotuline{\requirement{3}{inline-struct} defining a struct type inline}.
An additional syntax shown in the second expression
allows constructing a struct where the types and optional names
are defined by the expressions of the field values.
If no name is provided for a particular field,
that field cannot be accessed;
however, since the whole struct can be coerced
into another struct type with compatible field types
(such as in a function call as we discuss in more detail below),
such \dotuline{\requirement{3}{anonymous-struct} ``anonymous'' structs}
are still useful and very concise.
Presto and Athena also implement the \mintinline{SQL}{ROW} type.
Anonymous rows can be coerced into named ones as well;
however, the only mechanism to create named rows
is with a \mintinline{SQL}{CAST} expression
as shown in Listing~\ref{alg:struct:presto},
which is more verbose than BigQuery's inline declaration.
In Presto, fields of anonymous rows can be accessed with their ordinal index;
in Athena, fields of anonymous rows cannot be accessed at all.
Postgres supports anonymous structs with the \mintinline{SQL}{ROW} type;
however, only a subset of the expressions and functions on arrays
can deal with them.
For example, the \mintinline{SQL}{ARRAY_AGG} function
accepts anonymous structs as input
while the \mintinline{SQL}{ARRAY_CAT} function does not.
It is thus often necessary to cast anonymous rows to (previously created)
\dotuline{\requirement{3}{udts} user-defined types}
as shown in Listing~\ref{alg:struct:postgres}---%
a possibity that otherwise only SQL++ provides.

In JSONiq and SQL++, objects can be created with the \texttt{\{...\}} operator
containing any number of pairs of unique field names and values.
In SQL++, the rows produced by a nested \mintinline{SQL}{SELECT} statement
are of the same object types,
which thus offers an additional way to create (collections of) objects.
No anonymous objects exist;
in SQL++, columns without aliases are given generic names.

In all considered query languages, fields can be accessed
with the \texttt{.} operator known from many other programming languages,
with the particularity in Postgres that
``you often have to use parentheses to keep from confusing the parser.''%
~\cite{PostgresDocParsingStructs}

\vspace{-2ex}
\subsection{Accessing and Creating Nested Arrays}

\begin{flisting}
  \centering
  \begin{subflisting}{\linewidth}
    \begin{minted}{c++}
df.Define("goodJet_pt", "Jet_pt[abs(Jet_eta) < 1]")
    \end{minted}
    \vspace{-1ex}
    \caption{RDataFrames.}
    \label{alg:query3:rdf}
    \vspace{1.5ex}
  \end{subflisting}
  \begin{subflisting}{\linewidth}
    \begin{minted}{SQL}
SELECT j.pt FROM events
CROSS JOIN UNNEST(Jets) AS j
WHERE j.eta < 1
    \end{minted}
    \vspace{-1ex}
    \caption{BigQuery/Presto/Athena/Postgres.}
    \label{alg:query3:sql}
    \vspace{1.5ex}
  \end{subflisting}
  \begin{subflisting}{\linewidth}
    \begin{minted}{xquery}
$events.jets[][$$.eta < 1].pt
    \end{minted}
    \vspace{-1.5ex}
    \subcaption{JSONiq.}
    \label{alg:query3:jsoniq}
    \vspace{1.5ex}
  \end{subflisting}
  \begin{subflisting}{\linewidth}
    \begin{minted}{SQL}
SELECT VALUE j.pt
FROM events AS e, e.Jets AS j
WHERE ABS(j.eta) < 1
    \end{minted}
    \vspace{-1.5ex}
    \caption{SQL++.}
    \label{alg:query3:sqlpp}
  \end{subflisting}
  \vspace{-3ex}
  \caption{Simple unnesting of array elements.}
  \label{alg:query3}
\end{flisting}

Dealing with nested arrays is more involved
and can have various degrees of complexity.
We use Query~\queryref{3} as an example:
it filters the elements of a nested array (of structs),
flattens them,
and projects to one of the fields of the resulting structs.
In RDataFrames, we can assemble this query
from a large library of vectorized operations
that work on nested arrays directly
as shown in Listing~\ref{alg:query3:rdf}.
These operations include mathematical functions
like \mintinline{c++}{abs()},
Boolean operations like \mintinline{c++}{<},
and array selections based on bit-vectors like \mintinline{c++}{[.]},
as well as many others.
Unnesting for the purpose of aggregating across events is done implicitly.

In order to access array elements in SQL,
\mintinline{SQL}{UNNEST(.)} was introduced in SQL:1999.
Listing~\ref{alg:query3:sql} shows
how \dotuline{\requirement{1}{unnest} \mintinline{SQL}{UNNEST(.)} is
combined with \mintinline{SQL}{CROSS JOIN}},
which produces one row for each element of the provided array attribute
where all other attributes are duplicated.
The remainder of the query can thus filter and project
on any of the attributes as usual.
Note that this behavior is not strictly functional
and thus somewhat hard to understand:
\mintinline{SQL}{UNNEST(.)} on one attribute affects
the number of occurrences of the other attributes.%
\footnote{One could say that \mintinline{SQL}{UNNEST(.)}
  involves a ``spooky action at a distance''---a term
  that Einstein coined to express that, in quantum physics,
  choices made at a location seem to have an ``effect''
  faster than the speed of light
  on the outcome of measurements at remote locations.}
SQL++ also supports the \mintinline{SQL}{UNNEST(.)} construct,
though without the \mintinline{SQL}{CROSS JOIN} keyword,
and offers a short version with implicit unnesting
as shown in Listing~\ref{alg:query3:sqlpp}.

In JSONiq, such a simple query can use concise operators
for dealing with arrays, objects, and sequences thereof
(all of which are strictly functional):
The initial \mintinline{xquery}{.jets} extracts the \texttt{jets}
member of each object contained in the input
and produces a flat sequence of arrays;
the subsequent \texttt{[]} operator extracts elements of each of these arrays
and produces again a flat sequence of objects;
the predicate expression \texttt{[...]} then applies a filter
(using the context item \mintinline{xquery}{$$});
and the final \mintinline{xquery}{.pt} extracts a flat sequence of numbers.

\begin{flisting}
  \centering
  \begin{subflisting}{\linewidth}
    \begin{minted}{c++}
df.Filter("Sum(Jet_pt > 40) > 1")
    \end{minted}
    \vspace{-1.5ex}
    \caption{RDataFrames.}
    \label{alg:query4:rdf}
    \vspace{1.5ex}
  \end{subflisting}
  \begin{subflisting}{\linewidth}
    \begin{minted}{SQL}
... WHERE (SELECT COUNT(*)
           FROM UNNEST(events.Jets) AS j
           WHERE j.pt > 40) > 1
    \end{minted}
    \vspace{-1.5ex}
    \caption{BigQuery/Postgres with nested sub-query.}
    \label{alg:query4:bigquery}
    \vspace{1.5ex}
  \end{subflisting}
  \begin{subflisting}{\linewidth}
    \begin{minted}{SQL}
SELECT event_id, MET.sumet FROM events
CROSS JOIN UNNEST(events.Jets) AS j
WHERE j.pt > 40
GROUP BY event_id, MET.sumet
HAVING COUNT(*) > 1
    \end{minted}
    \vspace{-1ex}
    \subcaption{Presto/Athena/Postgres using \mintinline{SQL}{CROSS JOIN}.}
    \label{alg:query4:presto-crossjoin}
    \vspace{1.5ex}
  \end{subflisting}
  \begin{subflisting}{\linewidth}
    \begin{minted}{SQL}
... WHERE
  CARDINALITY(FILTER(events.Jets, j -> j.pt > 40)) > 1
    \end{minted}
    \vspace{-1ex}
    \subcaption{Presto/Athena using array functions.}
    \label{alg:query4:presto-arrayfunc}
    \vspace{1.5ex}
  \end{subflisting}
  \begin{subflisting}{\linewidth}
    \begin{minted}{xquery}
for $event in $events
where count($event.jets[][$$.pt > 40]) > 1
...
    \end{minted}
    \vspace{-2.5ex}
    \subcaption{JSONiq.}
    \label{alg:query4:jsoniq}
    \vspace{1.5ex}
  \end{subflisting}
  \begin{subflisting}{\linewidth}
    \begin{minted}{SQL}
... WHERE ARRAY_LENGTH(
  (SELECT * FROM e.Jets AS j WHERE j.pt > 40)) > 1)
    \end{minted}
    \vspace{-1ex}
    \subcaption{SQL++.}
    \label{alg:query4:sqlpp}
  \end{subflisting}
  \vspace{-3ex}
  \caption{Querying unnested array elements.}
  \label{alg:query4}
\end{flisting}

However, most real-world HEP queries are far more complex,
and the \mintinline{SQL}{CROSS JOIN} approach in SQL
does hence not seem to be a great fit.
Consider the only slightly more complex \queryref{4},
which filters events having at least two jets matching some predicate.
As Listing~\ref{alg:query4:presto-crossjoin} shows,
we can implement this query with a \mintinline{SQL}{CROSS JOIN},
which unnests the jets, filters them with the predicate,
and then counts the matching jets per event.
However, since the jets of all events are now flattened,
the latter operation requires a \mintinline{SQL}{GROUP BY},
which essentially undoes the flattening again.
Apart from potentially leading to a sub-optimal query plan,%
\footnote{\emph{If} \texttt{event\_id} is unique in the input,
  the unnesting produces runs of rows belonging to the same group,
  so a single-pass aggregation can be used.
  However, if the input is an external table
  made from files in Parquet or ROOT format,
  that information is not available.}
this makes both reading and writing the query
less natural than necessary.
Furthermore, this only works for a single array---%
a query that should filter or transform the elements of two or more arrays
has to be expressed as the join of two sub-queries
or as the sequence of two common table expressions,
which we discuss in more detail below.
Listing~\ref{alg:query4:presto-arrayfunc} shows
an alternative formulation based on the (non-standard)
\dotuline{\requirement{3}{array-function} array functions}
\mintinline{SQL}{CARDINALITY} and \mintinline{SQL}{FILTER}.
While this overcomes the aforementioned problems,
it does not work for the more complex patterns we discuss below.

Consider Listing~\ref{alg:query4:bigquery} as a contrast,
which shows the alternative of using
a \dotuline{\requirement{2}{nested-subquery} nested subquery},
which is part of SQL:1999.
In terms of semantics, the subquery can be thought of
to run once per row of the outer query,
providing the outer row as a constant.
Other than that, the two levels of the nesting
do not influence each other;
in particular, the subquery does not change the number of rows
of the outer query (unlike a \mintinline{SQL}{CROSS JOIN}).
Also, the subquery uses the same language constructs as the top-level query:
\mintinline{SQL}{UNNEST(events.Jets)} produces a table expression
that the remainder of the query uses as if it were a base table.
While BigQuery, Postgres, and (with a slightly different syntax) AsterixDB
support this construct, Athena and Presto do not.
Since in SQL++ tables and arrays are both collections,
which can be used as an input for array functions,
SQL++ offers some more freedom to combine the two paradigms
as shown in Listing~\ref{alg:query4:sqlpp}.

JSONiq has similar constructs illustrated in Listing~\ref{alg:query4:jsoniq}:
The outer level uses a ``FLWOR'' expression,
roughly speaking a generalization of SQL's SELECT-FROM-WHERE
that has an imperative look-and-feel but is, in fact, declarative.
The \mintinline{xquery}{for} clause produces a stream of tuples
from the items in the input sequence \mintinline{xquery}{$events},
in each of which the item is ``bound'' to the given variable name.
Subsequent clauses modify this stream;
for example, the \mintinline{xquery}{where} clause
filters out tuples from the stream.
There is only one type of expression in JSONiq,
i.e., the result of any expression can potentially be used
as an input of any other expression.
In the example, this means we can use similar expressions
for unnesting and filtering of sequences as before
in order to express the predicate of the \mintinline{xquery}{where} clause.

With RDataFrames, more complex query logic that cannot be assembled
with the vectorized operations provided by the framework
has to be written as UDFs in C++.
The \mintinline{c++}{make_p4} function above is such an example.
Since the main query logic in the form of data frame operations
is written in C++ as well
and the framework runs both levels in the same (optimizing) C++ interpreter,
these UDFs are reasonably seamless and efficient.
However, as we illustrate in the remainder of this section,
large fractions of the query logic
end up being written in UDFs outside the RDataFrame API.

\begin{flisting}
  \centering
  \begin{subflisting}{\linewidth}
    \begin{minted}{SQL}
ARRAY(SELECT AS STRUCT ...) AS new_particle
    \end{minted}
    \vspace{-1ex}
    \caption{BigQuery.}
    \label{alg:arrayconstruction:bigquery}
    \vspace{1.5ex}
  \end{subflisting}
  \begin{subflisting}{\linewidth}
    \begin{minted}{SQL}
ARRAY(SELECT ...) AS new_particle
    \end{minted}
    \vspace{-1ex}
    \caption{Postgres.}
    \label{alg:arrayconstruction:postgres}
    \vspace{1.5ex}
  \end{subflisting}
  \begin{subflisting}{\linewidth}
    \begin{minted}{SQL}
SELECT event_id, ARRAY_AGG(...) AS new_particle
... GROUP BY event_id
    \end{minted}
    \vspace{-1.5ex}
    \subcaption{Presto/Athena.}
    \label{alg:arrayconstruction:presto}
    \vspace{1.5ex}
  \end{subflisting}
  \begin{subflisting}{\linewidth}
    \begin{minted}{xquery}
[for $event in $events ...]
    \end{minted}
    \vspace{-1.5ex}
    \subcaption{JSONiq.}
    \label{alg:arrayconstruction:jsoniq}
    \vspace{1.5ex}
  \end{subflisting}
  \begin{subflisting}{\linewidth}
    \begin{minted}{SQL}
SELECT (SELECT ... FROM ...) AS new_particle
    \end{minted}
    \vspace{-1ex}
    \subcaption{SQL++.}
    \label{alg:arrayconstruction:sqlpp}
  \end{subflisting}
  \vspace{-3ex}
  \caption{Creating arrays.}
  \label{alg:arrayconstruction}
\end{flisting}

An interesting variation of this pattern
is to \dotuline{\requirement{3}{array-construction} produce a new array}
of particles in each event,
typically derived from the existing particles in that event.
While none of the queries in the benchmark explicitly ask to do that,
we have found it to be extremely useful for debugging
as well as for assembling complex chains
of transformation and filter stages (see below).
Listing~\ref{alg:arrayconstruction}
shows this pattern in the query languages:
In BigQuery, Postgres, and SQL++,
arrays can be constructed from any subquery
(with minor syntactic differences),
whereas Presto and Athena require the
\mintinline{SQL}{UNNEST}/\mintinline{SQL}{GROUP BY} pattern
using \mintinline{SQL}{ARRAY_AGG}.
In JSONiq, the items returned by any expression
can be turned into an array using the \texttt{[]} expression.
With RDataFrames, UDFs can create arrays by returning
an instance of the generic \mintinline{c++}{ROOT::RVec} type.

\subsection{Particle Combinations}
\label{sec:implementation:combinations}

\begin{flisting}
  \centering
  \begin{subflisting}{\linewidth}
    \begin{minted}{c++}
df.Define("indices", "Combinations(Jet_p4, 2)")
    \end{minted}
    \vspace{-1ex}
    \caption{RDataFrames.}
    \label{alg:particle-combinations:rdf}
    \vspace{1.5ex}
  \end{subflisting}
  \begin{subflisting}{\linewidth}
    \begin{minted}{SQL}
... FROM UNNEST(events.Jets) j1 WITH OFFSET i,
         UNNEST(events.Jets) j2 WITH OFFSET j
    WHERE i < j ...
    \end{minted}
    \vspace{-2ex}
    \caption{BigQuery.}
    \label{alg:particle-combinations:bigquery}
    \vspace{1.5ex}
  \end{subflisting}
  \begin{subflisting}{\linewidth}
    \begin{minted}{SQL}
... FROM UNNEST(Jets) WITH ORDINALITY AS j1,
         UNNEST(Jets) WITH ORDINALITY AS j2
    WHERE j1.ordinality < j2.ordinality ...
    \end{minted}
    \vspace{-1ex}
    \caption{Postgres.}
    \label{alg:particle-combinations:postgres}
    \vspace{1.5ex}
  \end{subflisting}
  \begin{subflisting}{\linewidth}
    \begin{minted}{SQL}
SELECT j1, j2, ... FROM events
CROSS JOIN
    UNNEST(Jets) WITH ORDINALITY AS j1(..., idx)
CROSS JOIN
    UNNEST(Jets) WITH ORDINALITY AS j2(..., idx)
WHERE j1.idx < j2.idx ...
    \end{minted}
    \vspace{-1.5ex}
    \caption{Presto.}
    \label{alg:implementation:sql:presto:cross-join}
    \vspace{1.5ex}
  \end{subflisting}
  \begin{subflisting}{\linewidth}
    \begin{minted}{xquery}
... (for $jet1 at $i in $event.jets[]
     for $jet2 at $j in $event.jets[]
     where $i < $j ...) ...
    \end{minted}
    \vspace{-1.5ex}
    \subcaption{JSONiq.}
    \label{alg:implementation:jsoniq:for-loop}
    \vspace{1.5ex}
  \end{subflisting}
  \begin{subflisting}{\linewidth}
    \begin{minted}{SQL}
... FROM e.Muon AS m1 AT idx1,
         e.Muon AS m2 AT idx2
WHERE idx1 < idx2 ...
    \end{minted}
    \vspace{-1ex}
    \subcaption{SQL++.}
    \label{alg:particle-combinations:sqlpp}
  \end{subflisting}
  \vspace{-3ex}
  \caption{Creating combinations of two jets.}
  \label{alg:particle-combinations}
\end{flisting}

While the patterns above deal with individual particles
(or a certain number thereof),
most real-world HEP queries actually involve
\emph{combinations} of particles.
Combinations may be
\dotuline{\requirement{1}{asym-combination} \emph{asymmetric}}
(such as in ``any electron-muon pair'')
or \dotuline{\requirement{1}{sym-combination} \emph{symmetric}}
(such as in ``any three jets'')
and may consist of combinations of two or more particles.
Since only particle combinations
within the same event are interesting,
it is natural to think of one event in isolation.
With that perspective,
asymmetric combinations are simple Cartesian products,
while symmetric combinations are only
a ``diagonal half'' of such products,
i.e., exactly one of $(p_i,p_j)$ and $(p_j,p_i)$
is in the result and $(p_i,p_i)$ is not.

We illustrate how the various programming interfaces express
this pattern in Listing~\ref{alg:particle-combinations}.
The RDataFrames API provides the vectorized
\mintinline{c++}{Combinations} operation for that purpose
(Listing~\ref{alg:particle-combinations:rdf}),
which produces indices of the desired combinations into the input array,
which can in turn be used to access the actual particles.
For use inside UDFs, the ROOT framework provides
the semantically equivalent \mintinline{c++}{VecOps::Combinations} function.

For the query languages, the patterns are similar as before:
SQL subqueries can simply use several calls to \mintinline{SQL}{UNNEST}
in their \mintinline{SQL}{FROM} clause,
which produces the Cartesian product of their content as usual.
In bot hcases, combining unnested particles also works outside of sub-queries,
however, with the same drawbacks as discussed above.
Duplicates can be eliminated with a filter on the indices,
which are produced by the \mintinline{SQL}{WITH OFFSET} clause
in BigQuery as shown in Listing~\ref{alg:particle-combinations:bigquery}
and by the \mintinline{SQL}{WITH ORDINALITY} clause
in Postgres (Listing~\ref{alg:particle-combinations:postgres}).
SQL++'s equivalent, the \mintinline{SQL}{AT} clause
shown in Listing~\ref{alg:particle-combinations:sqlpp}, is even more concise,
though currently not officially supported and thus undocumented.
We still use it in our study
because it produces correct results in the queries of the benchmark
and is expected to be completed soon~\cite{AsterixDBMailingListPosVar2021}.
Presto follows the syntax from the standard,
which enumerates the full list of fields names for specifying an alias.
Athena, BigQuery, and Postgres, in contrast,
allow to give an \dotuline{\requirement{3}{unnest-struct} alias
for the struct \emph{as a whole}}
reducing the verbosity significantly.
Postsgres supports both versions,
SQL++ only the short one
(but its name resolution scheme allows to access the struct's fields
without specifying the alias of the struct in some situations).

In Presto, the alternative of using array functions
extends to symmetric combinations
thanks to its array function \mintinline{SQL}{COMBINATIONS};
however, it is non-standard,
does not (easily) work for asymmetric combinations,
and is not implemented by any other system we are aware of,
so we do not discuss this alternative further.
Interestingly, this function (like a few other array functions)
is not supported by Athena
even though it originates from the same code base.

In JSONiq, the FLWOR expression may contain
several \mintinline{xquery}{for} clauses,
which essentially produce the Cartesian product
of their input sequences like in SQL.
With \mintinline{xquery}{at $i}, indices can be produced
and, of course, expressions can be nested arbitrarily as before.
In all query languages, creating arrays is done as before
(via \mintinline{SQL}{ARRAY(.)}, \mintinline{SQL}{ARRAY_AGG},
and the \texttt{[...]} operator, respectively).

\subsection{Multiple Transformations and Filters}

Most HEP queries do not consist
of a single transformation or filter using the patterns above,
but a \emph{series} of them:
physicists typically first compute some basic properties of each event,
which they may use in an initial filter,
and then iteratively add more of them to refine their search,
often reusing properties of the previous steps.
With RDataFrames, the user can specify a sequence
(or even a tree, if they specify more than one sink)
of transformations and filters, which can be chained arbitrarily.

In contrast, due to its lack of
\dotuline{\requirement{2}{variables} ``variables,''}
the same task is somewhat cumbersome in SQL---%
it is not possible to define a column alias with \mintinline{SQL}{AS}
and use that alias for the computation of other columns.
For example, \queryref{6} and \queryref{8}
combine several particles into one pseudo-particle,
which consists of a vector space transformation, a piece-wise addition,
and a reverse vector space transformation.
If we want to use this pseudo-particle more than once,
for example, to filter on one property and plot another one,
we need to spell out the computation of the pseudo-particle \emph{repeatedly}.

\begin{flisting}
  \centering
  \begin{subflisting}{\linewidth}
    \begin{minted}{SQL}
WITH Leptons AS (...),
TriLeptonsWithOtherLepton AS (
    SELECT *, (...) AS BestTriL
    FROM Leptons AS l
    WHERE ARRAY_LENGTH(l.Leptons) >= 3),
TriLeptonsWithMassAndOtherLepton AS (
    SELECT *, TrMass(MET, BestTriL.other) AS trMass
    FROM TriLeptonsWithOtherLepton
    WHERE BestTriLepton IS NOT NULL)
...
    \end{minted}
    \vspace{-2ex}
    \caption{BigQuery.}
    \label{alg:sequences:bigquery}
    \vspace{1.5ex}
  \end{subflisting}
  \begin{subflisting}{\linewidth}
    \begin{minted}{xquery}
for $event in parquet-file($input-path)
let $leptons := hep:concat-leptons($event)
let $best-tri-lepton := (...)
where exists($best-tri-lepton)
let $other := (...)
let $trMass := hep:TrMass($event.MET, $other)
...
    \end{minted}
    \vspace{-2ex}
    \subcaption{JSONiq.}
    \label{alg:sequences:jsoniq}
  \end{subflisting}
  \vspace{-3ex}
  \caption{Sequence of transformations in \queryref{8} (simplified).}
  \label{alg:sequences}
\end{flisting}

In the SQL implementations of the benchmark,
we found that a sequence of common table expressions or CTEs
(i.e., \mintinline{SQL}{WITH} statements)
resulted in the most concise and readable code.
Listing~\ref{alg:sequences:bigquery} shows an example:
each CTE adds one or more attributes
such as \texttt{BestTriL} and \texttt{trMass}
and passes through all existing ones (with \texttt{*})
such that the subsequent CTEs
or the final \mintinline{SQL}{SELECT} statement can use them.
The other SQL dialects work the same way.
While this is reasonably concise and avoids repeated code,
it involves the \emph{outer} level of relations
even though the computations only concern the \emph{inner} level of events.
Furthermore, the necessary reference to the previous CTE(s)
in the \mintinline{SQL}{WITH} clause
is more verbose than necessary.

A special case of lack of variables in SQL
is the fact that the standard does not allow
to \dotuline{\requirement{2}{variables-grouping} use a column alias
in the \mintinline{SQL}{GROUP BY} clause}.
Since \emph{all} queries consist of computing histograms,
i.e., counting occurrences of values \emph{per bin},
this is a frequent pattern.
In the dialects where this is the case,
we thus use at least one CTE to derive the desired property
and a final \mintinline{SQL}{SELECT} statement just for the histogram
(otherwise, the computation of the property
would have to be repeated in the \mintinline{SQL}{GROUP BY} clause).
Since BigQuery and Postgres \emph{do} allow one
to use aliases in the situation at hand,
they allow for more concise queries than the other SQL systems
(but diverge from the standard in this respect).

In JSONiq, variables are an integral part
of the pseudo"-imperative programming model of its FLWOR expression.
Since in that expression, \mintinline{xquery}{for}, \mintinline{xquery}{let},
\mintinline{xquery}{where}, \mintinline{xquery}{return},
and a few other clauses can be chained in essentially arbitrary order,
it is possible to assemble sequences of transformations and filters
in a single, top-level FLOWR expression
(which potentially uses nested FLOWR expressions on nested data).
Listing~\ref{alg:sequences:jsoniq} shows the same example as earlier.
Each \mintinline{xquery}{let} clause introduces a new variable
that any of the subsequent clauses and nested expressions can use---%
a concept that not only results in more concise code
but is also familiar to programmers of virtually any background.
SQL++ also has a \mintinline{SQL}{LET} clause
that essentially solves the problems of SQL discussed above;
however, it can be used somewhat less freely than that of JSONiq.

\subsection{User-defined Functions}

In all of the above, we give examples of query logic
that may be reused both within the same query and across them,
so a means to \dotuline{\requirement{1}{udfs} encapsulate query logic}
in user-defined functions (UDFs) or similar is essential.
An example for the first type of reuse is \queryref{7},
which asks for a ``lepton'' with a particular property,
i.e., either an electron or muon with that property
(these two particle types both being leptons).
Since electrons and muons of an event are stored in two different columns,
we either need to repeat the computation of the desired property
or encapsulate it as a (temporary) UDF.
With RDataFrames, the user can write UDFs in C++
for that purpose as described above.
The second type of reuse is more common and much more important:
the ROOT framework and similar tools for HEP analyses
come with a large collection of what could be called ``business logic,''
which encapsulate the computations of physical properties
such as a certain derived property of a particle,
the combination of several particles into one pseudo-particle, etc.
These computations often include mathematical formulae
and spelling them out for every query
would be tedious and error-prone.

The support for UDFs in the SQL-based systems in our study is mixed.
Athena does not support any type of UDFs suitable for our use case.
The offered UDFs are based on serverless functions
and thus need to ship all data to a different cloud service,
invoke the UDF for each record at the time,
incur further costs, and are subject to concurrency quotas---%
in short, they are not suitable for data-intensive tasks.
Presto recently added experimental support for UDFs---%
currently with the severe limitation
that UDFs cannot call other UDFs,
making it impractical to implement real-world function libraries.
However, given the current development effort on this feature,
this limitation is likely to be lifted soon.
BigQuery, Postgres, and SQL++ have mature support
for both permanent and temporary UDFs.
JSONiq allows declaring functions as part of the query text
and to import functions and constants from external modules.
Since the full name of such a module is a URI
that typically hosts the code of that module publicly on the web,
this mechanism can be seen as a simple built-in package manager.

\begin{flisting}
  \centering
  \begin{subflisting}{\linewidth}
    \begin{minted}{SQL}
CREATE TEMP FUNCTION AddPtEtaPhiM2(
   pepm1 STRUCT<Pt FLOAT64, Eta FLOAT64,
                Phi FLOAT64, Mass FLOAT64>,
   pepm2 STRUCT<Pt FLOAT64, Eta FLOAT64,
                Phi FLOAT64, Mass FLOAT64>) AS ...
SELECT AddPtEtaPhiM2(
     STRUCT(l1.Pt, l1.Eta, l1.Phi, l1.Mass),
     STRUCT(l2.Pt, l2.Eta, l2.Phi, l2.Mass)), ...
FROM UNNEST(Leptons) l1, UNNEST(Leptons) l2 ...
    \end{minted}
    \vspace{-1ex}
    \caption{BigQuery.}
    \label{alg:udfs:bigquery}
    \vspace{1.5ex}
  \end{subflisting}
  \begin{subflisting}{\linewidth}
    \begin{minted}{SQL}
CREATE FUNCTION AddPtEtaPhiM2(
    IN pepm1 anyelement, IN pepm2 anyelement) ...
SELECT AddPtEtaPhiM2(l1, l2)
FROM UNNEST(Leptons) l1, UNNEST(Leptons) l2 ...
    \end{minted}
    \vspace{-1ex}
    \caption{Postgres.}
    \label{alg:udfs:postgres}
    \vspace{1ex}
  \end{subflisting}
  \begin{subflisting}{\linewidth}
    \begin{minted}{xquery}
declare function hep:add-PtEtaPhiM($p1, $p2) { ... };
for $l1 at $i in $leptons
for $l2 at $j in $leptons
let $mass := hep:add-PtEtaPhiM($l1, $l2).mass
...
    \end{minted}
    \vspace{-2ex}
    \subcaption{JSONiq.}
    \label{alg:udfs:jsoniq}
    \vspace{1.5ex}
  \end{subflisting}
  \begin{subflisting}{\linewidth}
    \begin{minted}{SQL}
DECLARE FUNCTION AddPtEtaPhiM2(p1, p2) {...};
FROM Leptons AS l1 AT idx1, Leptons AS l2 AT idx2
SELECT AddPtEtaPhiM2(l1, l2).mass ...
    \end{minted}
    \vspace{-1ex}
    \subcaption{SQL++.}
    \label{alg:udfs:sqlpp}
  \end{subflisting}
  \vspace{-2.5ex}
  \caption{Declaration and call of a UDF in \queryref{8} (simplified).}
  \label{alg:udfs}
\end{flisting}

UDFs in both Presto and BigQuery
\dotuline{\requirement{2}{udfs-structs} support structs as function parameters};
however, the field names and types of the structs
need to be specified in the declaration,
and the value used in the invocation
needs to have matching arity and compatible types.%
\footnote{In particular, it is not possible
  to access fields of a function argument
  declared as \mintinline{SQL}{ANY TYPE} in BigQuery.}
This allows
for providing anonymous structs as discussed above;
however, the invoking code needs to
(1) project away all fields of an existing \mintinline{SQL}{ROW} instance
that the function does not list in its parameter and
(2) bring the remaining ones in the same order.
Listing~\ref{alg:udfs:bigquery} illustrates
how this affects verbosity in BigQuery:
The full list of field names need to be specified twice,
once in function declaration and once for assembling the call value.
Listings~\ref{alg:udfs:postgres}, \ref{alg:udfs:jsoniq},
and \ref{alg:udfs:sqlpp} show the corresponding example
in Postgres' dialect, JSONiq, and SQL++, respectively.%
\footnote{Notice in the example
          that SQL++ allows to specify the \mintinline{SQL}{SELECT}
          as the \emph{last} clause of the statement.}
Both the function declaration and the invocation site
use objects without explicit enumeration of member names---%
the members accessed by the function body must be present
(otherwise, depending on the body, an error is raised)
and the superfluous members at the invocation are simply ignored
(and may be physically removed by an optimizer).
Furthermore, the order of the members does not matter
(as it is undefined).

\begin{flisting}
  \centering
  \begin{subflisting}{\linewidth}
    \begin{minted}{c++}
df.Histo1D({"histogram name", "title;x-label;y-label",
            100, 0, 2000}, "values");
    \end{minted}
    \vspace{-1.5ex}
    \caption{RDataFrames.}
    \label{alg:histogram:rdf}
    \vspace{1.5ex}
  \end{subflisting}
  \begin{subflisting}{\linewidth}
    \begin{minted}{SQL}
SELECT HistogramBin(value, 15, 250, 100) AS x,
       COUNT(*) AS y
FROM previousCTE GROUP BY x ORDER BY x
    \end{minted}
    \vspace{-1ex}
    \caption{BigQuery.}
    \label{alg:histogram:bigquery}
    \vspace{1.5ex}
  \end{subflisting}
  \begin{subflisting}{\linewidth}
    \begin{minted}{xquery}
hep:histogram($values, 15, 250, 100)
    \end{minted}
    \vspace{-1.5ex}
    \subcaption{JSONiq.}
    \label{alg:histogram:jsoniq}
    \vspace{1.5ex}
  \end{subflisting}
  \begin{subflisting}{\linewidth}
    \begin{minted}{SQL}
histogram((FROM ... SELECT ...), 15, 250, 100)
    \end{minted}
    \vspace{-1ex}
    \subcaption{SQL++.}
    \label{alg:histogram:sqlpp}
  \end{subflisting}
  \vspace{-3ex}
  \caption{Histogram computation.}
  \label{alg:histogram}
\end{flisting}

The SQL-based systems have additional limitations. They do not support
\dotuline{\requirement{2}{udfs-tables} UDFs that consume or produce a table}.
It is thus not possible to encapsulate
the computation of histograms fully.
Instead, the user needs to spell out
the grouping with aggregation manually for every query
as illustrated in Listing~\ref{alg:histogram:bigquery}.
In JSONiq and SQL++, functions can work on any level
so this logic can be fully hidden
as illustrated in Listings~\ref{alg:histogram:jsoniq}
and \ref{alg:histogram:sqlpp}.
Also with the RDataFrames API,
only the built-in transformations can be applied to data frames;
however, a large number of such transformations and sinks exists,
in particular, domain-specific ones
such as the \mintinline{c++}{Histo1D} sink
shown in Listing~\ref{alg:histogram:rdf},
which computes the aggregation required for the histogram
(and immediately produces the actual plot).
Furthermore, SQL dialects do not allow
declaring variables inside of UDFs,
requiring a series of helper functions
to avoid repeating common sub-expressions
(see the CTEs discussed above).
The UDFs in the SQL dialects studied
are not standard and mutually incompatible.

\subsection{Summary}

We summarize the discussion of this section
in Table~\ref{tbl:requirements-summary}.
In addition to marking unsupported features with a dash (-),
we also indicate ``how well'' a system supports the features it supports
(more asterisks indicating better support).
Again, this cannot be seen as a fully objective quantification
but rather as an approximate visualization of the detailed discussion above.
The table suggests that JSONiq and SQL++ are best suited for HEP analyses---%
since they were purpose-built for the JSON data model,
which is also heavily nested (and heterogeneous),
this is not completely surprising.
BigQuery's and Postgres' SQL dialects
implement all related features of the standard
and have a few proprietary extensions,
which together make them a good match as well;
the only major missing constructs
are the lack of variables and table-based UDFs.
In contrast, we believe that Athena cannot currently be considered viable
as the lack of UDFs makes it impossible to share library code between users.
While Presto is on the brink of having this feature,
it shares a number of missing or cumbersome constructs with its fork Athena
that make it a less-than-ideal (though viable) system for HEP.
RDataFrames are of course well suited for what they are built for;
however, the fact that they make the columnar storage format
part of the programming model requires a higher programming effort
than that of a suitable declarative query language.
Overall, we believe that the NF$^2$ support added with SQL:1999
as well as more modern languages like JSONiq and SQL++
have the potential to make general-purpose data processing systems
a viable alternative to the domain-specific systems used today.

\begin{table}[t]
  \caption{Summary of functionality of general-purpose
           data processing systems for HEP analyses.}%
  \label{tbl:requirements-summary}%
  \vspace{-3ex}
  \centering%
  \newcommand{\system}[1]{\rotatebox[origin=c]{90}{\textbf{#1}}}%
  \newcommand{\w}{\textasteriskcentered}%
  \newcommand{\ww}{\textasteriskcentered\textasteriskcentered}%
  \newcommand{\www}{\textasteriskcentered\textasteriskcentered\textasteriskcentered}%
  \resizebox{\linewidth}{!}{%
  \setlength{\tabcolsep}{2.5pt}%
  \begin{tabular}{@{}l@{\hspace*{0pt}}ccccccc@{}}
    \toprule
      & \system{Athena} & \system{BigQuery} & \system{Postgres} & \system{Presto} & \system{JSONiq} & \system{SQL++} & \system{RDataFrame} \\
    \midrule
    \requirementref{1}{unnest} unnest arrays                & \ww  & \ww  & \ww  & \ww  & \www & \www & \ww  \\
    \requirementref{1}{asym-combination} asym. combinations & \www & \www & \www & \ww  & \www & \www & \ww  \\
    \requirementref{1}{sym-combination} sym. combinations   & \www & \www & \www & \ww  & \www & \www & \ww  \\
    \requirementref{1}{udfs} UDFs                           & -    & \ww  & \ww  & \w   & \www & \www & \www \\
    \midrule
    \requirementref{2}{structs} structured types            & \ww  & \www & \ww  & \ww  & \www & \www & \ww  \\
    \requirementref{2}{nested-subquery} nested sub-query    & -    & \www & \www & -    & \www & \www & \ww  \\
    \requirementref{2}{variables} variables                 & -    & -    & -    & -    & \www & \ww  & \www \\
    \requirementref{2}{variables-grouping} group by variable& -    & \www & \www & -    & \www & \ww  & n/a  \\
    \requirementref{2}{udfs-structs} struct params in UDFs  & \w   & \ww  & \www & \ww  & \www & \www & \www \\
    \requirementref{2}{udfs-tables} tables in UDFs          & -    & -    & -    & -    & \www & \www & -    \\
    \midrule
    \requirementref{3}{inline-struct} inline struct types   & -    & \www & -    & -    & \www & \www & -    \\
    \requirementref{3}{anonymous-struct} anonymous structs  & \ww  & \www & \ww  & \www & -    & -    & -    \\
    \requirementref{3}{udts} user-defined types             & -    & -    & \ww  & -    & -    & \www & \www \\
    \requirementref{3}{array-function} array functions      & \ww  & \ww  & \ww  & \www & \ww  & \ww  & \ww  \\
    \requirementref{3}{array-construction} array construction&-    & \ww  & \ww  & -    & \www & \www & \ww  \\
    \requirementref{3}{unnest-struct} unnest whole structs  & \www & \www & \www & -    & \www & \www & -    \\
    \midrule
    \#characters                        & 6.7k   & 7.6k &  7.6k    & 7k & \textbf{3.8k}         &   \textbf{3.8k}   &  11k  	\\
    \#lines                             & 343    & 280           &  286    & 274  & \textbf{106} &  175     & 236 	\\
    \#clauses                           & 222   & 223           &  205    & 180  & \textbf{56}  & 104     & 134 	\\
    avg. \#clauses/query             	& 24.6   & 16 	     &  17    & 19 & \textbf{6.2} &    8.6  & 14.9 	\\
    \#unique clauses                    & 24    & 20            &  20    &  27  & \textbf{8}   & 16     & 15 	\\
    avg. \#unique clauses/query      	& 12.1  & 8.3           &  8.6    & 10  & \textbf{3.3} &  4.9    & 7 	\\
    \bottomrule				
  \end{tabular}%
  }%
  \vspace{-3.7ex}%
\end{table}

We have implemented the full benchmark
in the seven languages, dialects, and programming interfaces.
Table~\ref{tbl:requirements-summary} shows
the overall implementation length in various metrics:
the number of characters and lines
(which exclude white space, blank lines, and comments),
the number of clauses (to which we include calls to built-in functions),
the number of unique clauses per query, 
the number of unique clauses
(counting how many \emph{different} language constructs are used overall),
and the number of average unique clauses per query.
All six metrics show the same picture:
SQL++ and JSONiq are most concise; Athena,
Postgres, BigQuery and Presto require more code.
Such metrics always need to be interpreted with care:
the metrics vary with implementation style, formatting, etc.,
more concise queries are not always more readable,
and readability depends to a large degree
on the personal preference and experience of the programmer.
Still, the numbers do provide a useful quantification
of the previous discussion of this section
and suggest a similar conclusion.
We have also made the full implementations publicly available%
~\cite{BenchmarkScripts},
as a reference for the reader.

% The following section should contain the results of evaluation across the evaluation criteria
\section{Performance Evaluation}
\label{sec:evaluation}
We now study the efficiency and scalability
of the systems implementing the query languages
of the previous section.

\subsection{Experimental Setup}

\textbf{Platform.}~~%
For the self-managed systems,
we use virtual machines in Amazon EC2 from the \texttt{m5d} series.
The largest size of that instance type, \texttt{24xlarge},
has 48 real CPU cores, \SI{384}{\gibi\byte} of main memory,
four NVMe SSDs of \SI{900}{\gibi\byte} each configured as RAID 0,
and \SI{25}{\giga\bit\per\second} networking,
and costs \SI{6.048}{\$\per\hour} in the \texttt{eu-wests-1} region;
all numbers are proportionally smaller for the smaller sizes.

\textbf{Storage.}~~%
We use the original ROOT files for RDataFrames
and Parquet files for all other systems.
For AsterixDB, Athena, BigQuery, Presto, and RumbleDB,
we place them on cloud storage
and process directly from there.
For RDataFrames, we put the input files on the local disk,
which gives a similar performance
as the typically used \texttt{xrootd} network storage protocol,
which, in turn, is more performant than S3 for the required access patterns.
For Postgres, we use the experimental foreign-data wrapper
for Parquet files~\cite{ParquetFdw} and place the files on SSD
since it does not support network storage.

\textbf{Input Data Size.}~~%
We use the original data set defined by the benchmark~\cite{CmsOpenData2012}.
Additionally, we define a \emph{scale factor} (SF),
which we use for simulating more realistic data set sizes
of up to \SI{2}{\tera\byte}.
Even though the original files are from experiments from 2012,
they are the most up-to-date \emph{publicly available} files
due to the regular shut-down periods of the accelerator
and data retention policies of CERN. %
Since then, the amount of produced data
has grown exponentially~\cite{LuminosityPlot},
and data set sizes used by physicists today
as well as their expected size for the next decade,
are about one and two orders of magnitude larger%
~\cite{LhcAbout,Meglio2017,Calafiura2020}, respectively.
We thus replicate the original file SF times for $\text{SF} \ge 1$
and take the first fraction of SF events for  $\text{SF} \le 1$.

\begin{figure*}[t]
  \begin{subfigure}[t]{.245\linewidth}
    \includegraphics[scale=.7]{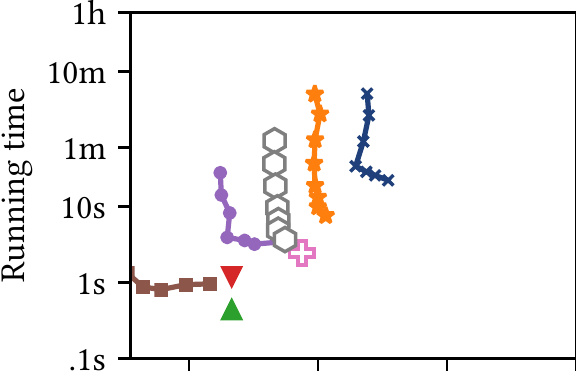}%
    \hfill
    \vspace{-1ex}
    \caption{\queryref{1}\hspace{-2em}~}
    \label{fig:cost-running-time-tradeoff:1}
  \end{subfigure}
  \begin{subfigure}[t]{.2\linewidth}
    \includegraphics[scale=.7]{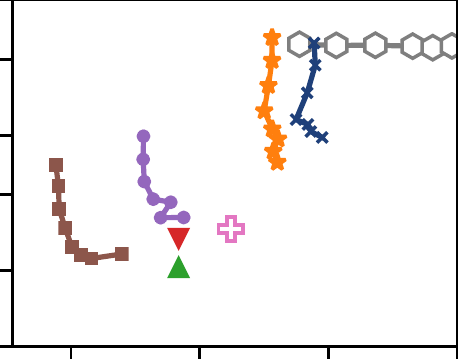}%
    \hfill
    \vspace{-1ex}
    \caption{\queryref{2}\hspace{1em}~}
    \label{fig:cost-running-time-tradeoff:2}
  \end{subfigure}
  \begin{subfigure}[t]{.2\linewidth}
    \includegraphics[scale=.7]{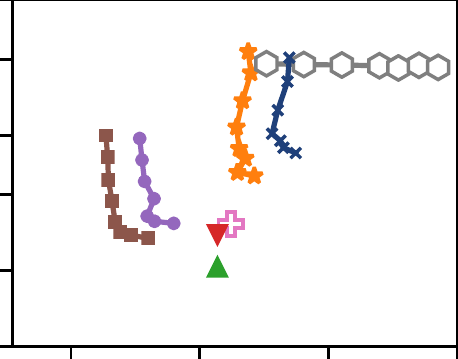}%
    \hfill
    \vspace{-1ex}
    \caption{\queryref{3}\hspace{1em}~}
    \label{fig:cost-running-time-tradeoff:3}
  \end{subfigure}
  \begin{subfigure}[t]{.31\linewidth}
    \tikz[overlay,remember picture] \node[anchor=south west,inner sep=0]
      (cost-running-time-tradeoff-4) {
        \includegraphics[scale=.7]{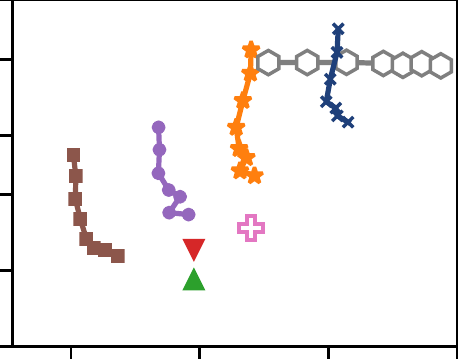}};%
    \hfill
    \vspace{-1ex}
    \caption{\queryref{4}\hspace{7em}~}
    \label{fig:cost-running-time-tradeoff:4}
  \end{subfigure}
  \begin{subfigure}[t]{.245\linewidth}
    \includegraphics[scale=.7]{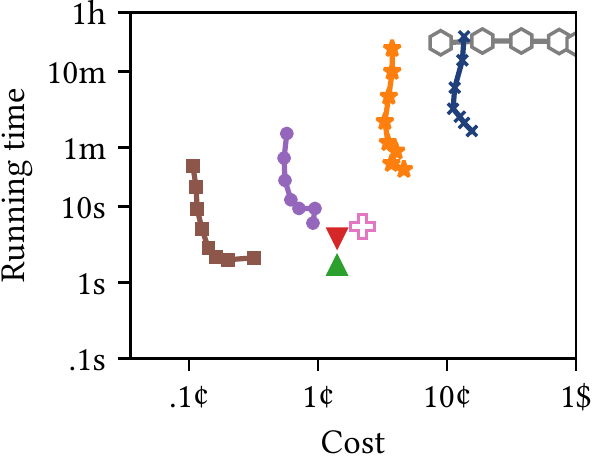}%
    \hfill
    \caption{\queryref{5}\hspace{-2em}~}
    \label{fig:cost-running-time-tradeoff:5}
  \end{subfigure}
  \begin{subfigure}[t]{.2\linewidth}
    \includegraphics[scale=.7]{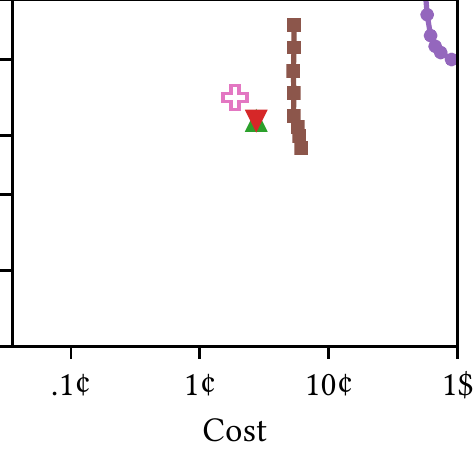}%
    \hfill
    \caption{\queryref{6a}\hspace{1em}~}
    \label{fig:cost-running-time-tradeoff:6-1}
  \end{subfigure}
  \begin{subfigure}[t]{.2\linewidth}
    \includegraphics[scale=.7]{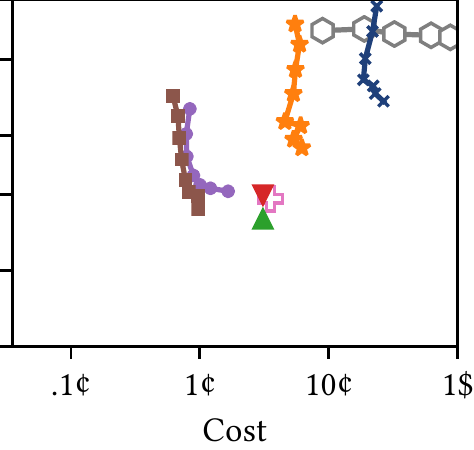}%
    \hfill
    \caption{\queryref{7}\hspace{1em}~}
    \label{fig:cost-running-time-tradeoff:7}
  \end{subfigure}
  \begin{subfigure}[t]{.31\linewidth}
    \tikz[overlay,remember picture] \node[anchor=south west,inner sep=0]
      (cost-running-time-tradeoff-8) {
        \includegraphics[scale=.7]{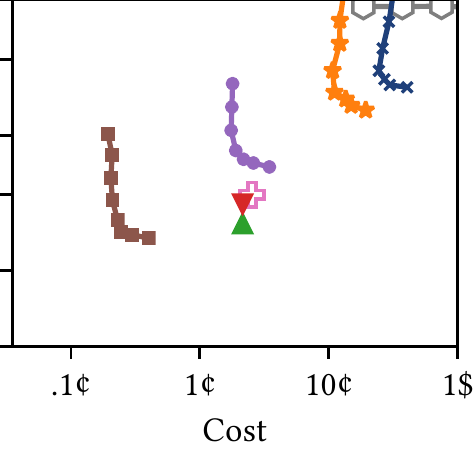}};%
    \hfill
    \caption{\queryref{8}\hspace{7em}~}
    \label{fig:cost-running-time-tradeoff:8}
  \end{subfigure}%
  \vspace{-2ex}%
  \caption{Running time/cost trade-off for various systems under test.}
  \label{fig:cost-running-time-tradeoff}
  \begin{tikzpicture}[overlay,remember picture]
    \node [fit=(cost-running-time-tradeoff-4)(cost-running-time-tradeoff-8),
           inner sep=0pt] (cost-running-time-tradeoff-right) {};
    \node [right=0em of cost-running-time-tradeoff-right,inner sep=0pt]
      (legend) {\includegraphics[scale=.7]{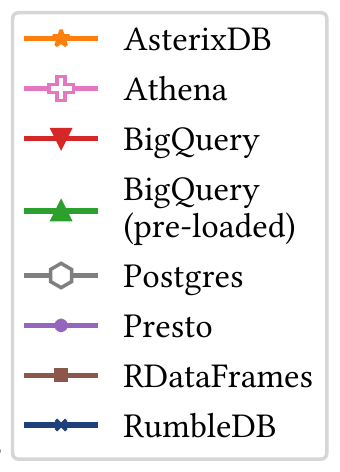}};
  \end{tikzpicture}
\end{figure*}

\subsection{End-to-End Comparison}

We start with a comparison among the systems under test
on their end-to-end performance and monetary cost.
For the self-managed systems,
we report numbers for all instance sizes
from \texttt{large} to \texttt{24xlarge}.
We compute the query cost
as the product of the number of wall-time seconds
and the per-second price of the underlying instance.
We use $\text{SF}=1$ with the exception of RumbleDB,
where we use the largest data set size
that the system can handle in at most \SI{10}{\minute}
and extrapolate from that number.
We report a single configuration for the QaaS systems
since we do not have any control over the amount of resources.

Figure~\ref{fig:cost-running-time-tradeoff} shows the result
for all queries except \queryref{6b},
which has nearly identical results as \queryref{6a}.
On all queries, BigQuery is the fastest system,
often even when using external tables,
answering all queries except \queryref{6} in less than \SI{10}{\second}
and many in a low single-digit number of seconds.
We discuss the special nature of \queryref{6} below.
Using pre-loaded data improves performance further
by about \SI{2}{\times} in most queries.
Athena is significantly slower:
it comes close to BigQuery in some queries (e.g., \queryref{3}),
but often has a significant margin,
though response time is still in the order
of \SIrange{10}{20}{\second} except for \queryref{6}.
The performance of the self-managed systems
generally depends on the instance size,
yielding lower running time using large instances but also a higher cost.
The only exception is Postgres:
despite extensive manual rewriting of the queries
and tuning of the configuration,
the system was only able to fully parallelize \queryref{1}---%
for all other queries, the system could not produce fully parallel plans
or did not execute the plans in parallel.
RDataFrames is a strong runner up; its fastest configuration
outperforms BigQuery with external tables for some queries
and only has a narrow gap on the other ones.
The good performance of RDataFrames can be explained
by its efficient, jit-compiled execution model.
Presto is again significantly slower, in particular, for small instance sizes,
though its fastest configuration comes close to Athena.
AsterixDB, Postgres, and RumbleDB, however,
are about one order of magnitude slower than the next system
and up to two orders of magnitude slower than the fastest.
Except for the most simple ones,
queries take minutes or even hours rather than seconds on these systems,
making them impractical for interactive analyses.
We attribute the slower performance of these systems
to their interpreted execution model;
in particular, AsterixDB and RumbleDB are designed
to work on heterogeneous data sets,
where polymorphic operators on polymorphic data representations
are hard to avoid.

\begin{figure*}[t]
  ~\hspace{1em}
  \begin{subfigure}[t]{.25\linewidth}
    \includegraphics[scale=.7]{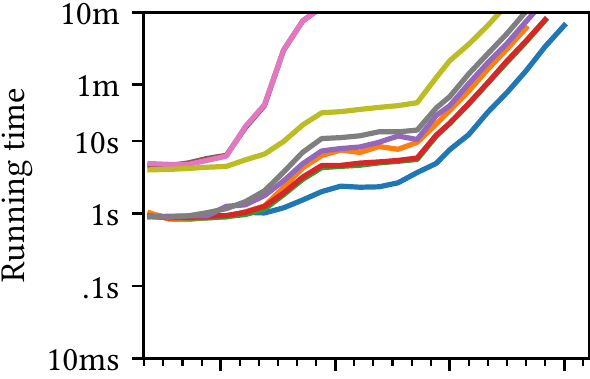}%
    \hfill
    \vspace{-1ex}
    \caption{AsterixDB\hspace{-3em}~}
    \label{fig:scaling-running-time:asterixdb}
  \end{subfigure}
  \begin{subfigure}[t]{.2\linewidth}
    \includegraphics[scale=.7]{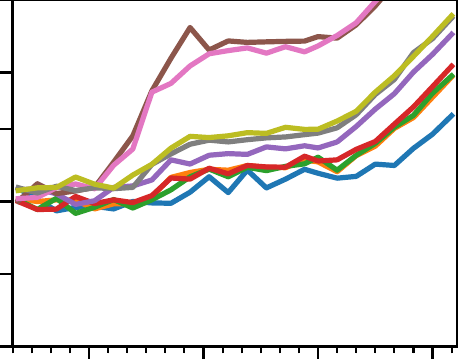}%
    \hfill
    \vspace{-1ex}
    \caption{Athena}
    \label{fig:scaling-running-time:athena-v2}
  \end{subfigure}
  \begin{subfigure}[t]{.2\linewidth}
    \includegraphics[scale=.7]{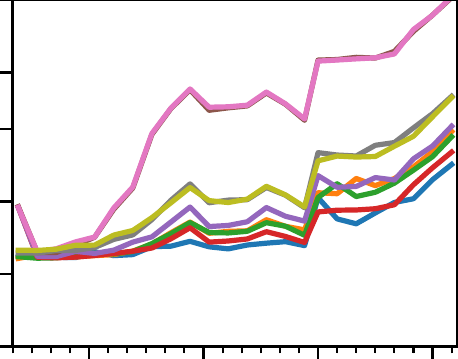}%
    \hfill
    \vspace{-1ex}
    \caption{BigQuery (pre-loaded)}
    \label{fig:scaling-running-time:bigquery}
  \end{subfigure}
  \begin{subfigure}[t]{.27\linewidth}
    \tikz[overlay,remember picture] \node[anchor=south west,inner sep=0]
      (scaling-running-time-top-right) {
        \includegraphics[scale=.7]
          {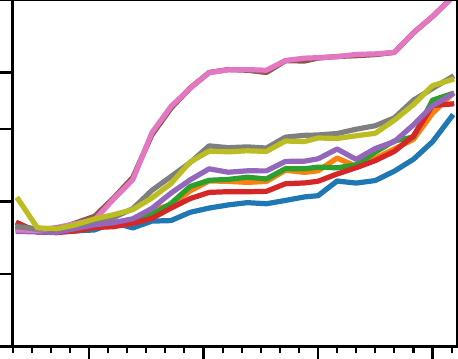}};%
    \hfill
    \vspace{-1ex}
    \caption{BigQuery\hspace{5em}~}
    \label{fig:scaling-running-time:bigquery-external}
  \end{subfigure}
  \hfill~\\
  ~\hspace{1em}
  \begin{subfigure}[t]{.25\linewidth}
    \includegraphics[scale=.7]{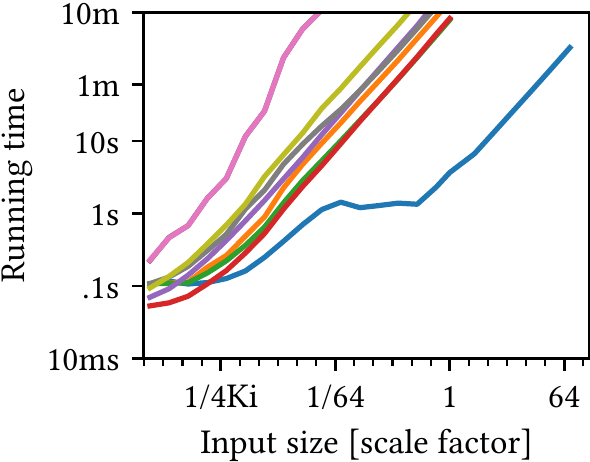}%
    \hfill
    \caption{Postgres\hspace{-3em}~}
    \label{fig:scaling-running-time:postgres}
  \end{subfigure}
  \begin{subfigure}[t]{.2\linewidth}
    \includegraphics[scale=.7]{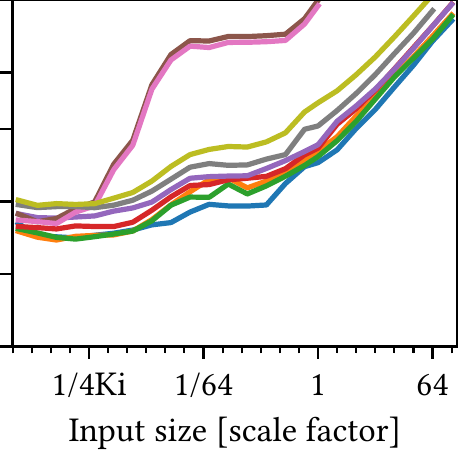}%
    \hfill
    \caption{Presto}
    \label{fig:scaling-running-time:presto}
  \end{subfigure}
  \begin{subfigure}[t]{.2\linewidth}
    \includegraphics[scale=.7]{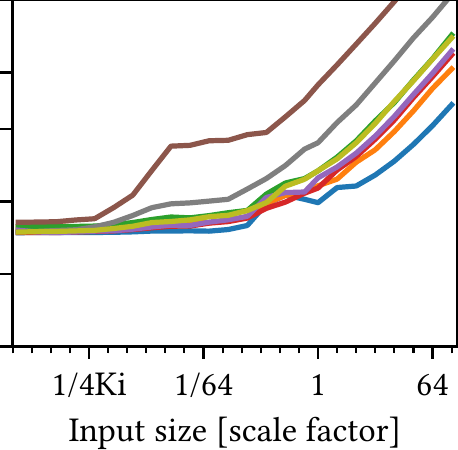}%
    \hfill
    \caption{RDataFrames}
    \label{fig:scaling-running-time:rdataframes}
  \end{subfigure}
  \begin{subfigure}[t]{.28\linewidth}
    \tikz[overlay,remember picture] \node[anchor=south west,inner sep=0]
      (scaling-running-time-bottom-right) {
        \includegraphics[scale=.7]{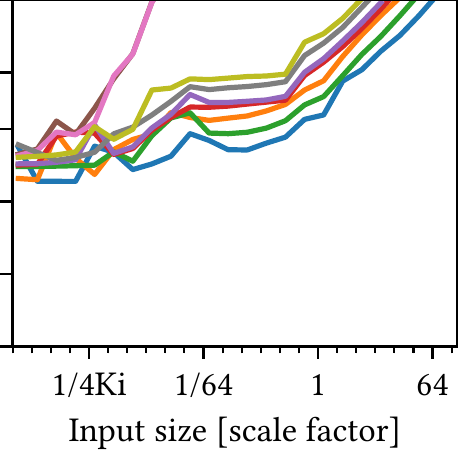}};%
    \hfill
    \caption{RumbleDB\hspace{6em}~}
    \label{fig:scaling-running-time:rumble}
  \end{subfigure}
  \hfill~
  \vspace{-3ex}
  \caption{Impact of data size on end-to-end running time
           of various systems under test.}
  \label{fig:scaling-running-time}
  \begin{tikzpicture}[overlay,remember picture]
    \node
      [fit=(scaling-running-time-top-right)(scaling-running-time-bottom-right),
       inner sep=0pt] (scaling-running-time-right) {};
    \node [right=0.5em of scaling-running-time-right,inner sep=0pt] (legend)
    {\includegraphics[scale=.7]{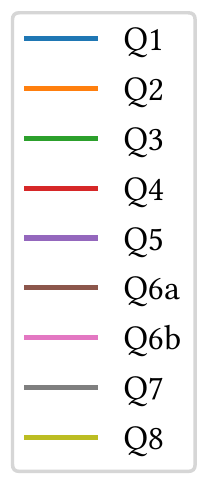}};
  \end{tikzpicture}
  \vspace{-2ex}
\end{figure*}

The monetary cost varies greatly among the systems as well.
Among the two QaaS systems, BigQuery is not only faster
but also cheaper for almost all queries.
This is due to an interplay of various factors.
Nominally, both systems charge \SI{5}{\$} for each TB scanned by the query;
however, the latter quantity is computed differently.
In BigQuery, it is the number of bytes in the full \emph{uncompressed} column;
i.e., for each column that is used by the query,
the size of the data type of that column
multiplied with the number of its entries.
Note that the system only exposes
double-precision floating-point numbers
(which count as \SI{8}{\byte}) to the user,
even if the underlying Parquet files in the external tables
actually store single-precision numbers (which only need \SI{4}{\byte}).
In Athena, only the bytes actually read from storage are taken into account,
i.e., the encoding and compression mechanisms of Parquet
have the potential to reduce the query price significantly.
However, most attributes in the data set consist of floating-point numbers,
so most columns in the Parquet files have only negligible compression ratios.
Furthermore, as we study in more detail below,
Athena (like Presto) is not able to push projections into structs,
i.e., if a query accesses only one or few fields of a struct
(at the top level or inside of array columns),
then still all columns of that struct are read from storage
and the query is charged accordingly.
Overall, this leads to a higher cost in Athena for almost all queries.
In particular, in \queryref{1}, \queryref{2}, and \queryref{4},
where only relatively few fields of large structs are accessed,
Athena's query price is higher due to the lack of push-down support.
In the remaining queries, relatively more fields are accessed
so the gap to BigQuery narrows;
in \queryref{6}, Athena is even slightly cheaper.

The cost of the self-managed systems
is linear on the queries' running time,
leading to a different picture:
For the computationally simple queries,
namely \queryref{1} to \queryref{5},
Presto and RDataFrames are significantly cheaper than the QaaS systems,
often by a factor \SIrange{2}{6}{\times}.
However, the gap narrows for \queryref{7} and \queryref{8},
which are more compute-intensive as indicated by the higher running times.
For \queryref{6}, Presto and RDataFrames
are more than one and about half an order of magnitude more expensive
than the QaaS systems, respectively.
This is mainly due to the pricing model of the latter,
in which computation is essentially free---%
the cloud provider presumably counts on
such compute-intensive queries being infrequent
and cross-finances them with the majority of scan-intensive ones.
AsterixDB, Postgres, and RumbleDB
are more expensive by the same factor they are slower:
at least an order of magnitude compared to the next system.
The potentially higher productivity of their query language
thus currently comes at a significantly higher monetary cost.

The numbers presented in this experiment
show only a partial picture of the total monetary cost.
First, using spot instances has the potential
to reduce the cost considerably, sometimes by up to \SI{5}{\times}.
Second, the user also has to pay
for the idle time of their instances,
which in turn could be reduced
with a multi-tenant cluster and/or auto-scaling.
The presented numbers, however, give \emph{some} indication
about how self-managed systems compare to QaaS systems.

\subsection{Scaling with Data Set Size}

We study the impact of different data set sizes.
We run the benchmark on $\text{SF} = 2^i$ for $i=-16,...,7$.
At the smallest scale factor, the input consists of about 800 events
and requires about \SI{250}{\kibi\byte};
at the largest scale factor, it consists of about \SI{6.8}{\giga\nothing} events
and requires slightly more than two terabytes.
For the self-managed systems,
we use the largest size (\texttt{m5d.24xlarge});
 cloud-based systems use the resources assigned by the cloud provider.
We exclude all configurations
that take longer than \SI{10}{\minute} to complete.

Figure~\ref{fig:scaling-running-time} shows the result.
The running time initially increases in all cases with the data size
and then reaches a plateau.
This is due to the granularity of parallelization:
All systems using Parquet files only parallelize
\emph{across} row groups, not \emph{within} them.
Each row group has an average of \SI{400}{\kilo\nothing} events,
which is the beginning of the plateau---%
execution is single-threaded for smaller data sets
and increases with the number of events,
and parallelized after that.
For Postgres, this is only true for \queryref{1}
due to its inability to parallelize the other queries
as discussed above.
The running time then increases again
at the end of the plateau.
This happens when there are more row groups than CPU cores:
at $\text{SF}=128$, the data set has in the order
of \SI{16}{\kilo\nothing} row groups in Parquet
while the instances have 48 real CPU cores
(i.e., 96 logical cores including SMT),
so the largest data sizes cannot be processed completely in parallel.
Interestingly, this is even true for the QaaS systems,
where our queries seem to exceed the amount of resources
that the cloud providers are willing to dedicate to a single query
for scale factors larger than about 2 and 8
for Athena and BigQuery, respectively.
At the largest scale factor,
all systems seem to reach a steady state
where the running time is dominated
by the raw processing throughput in terms of events per second.
Due to the structure of the queries
(a top-level aggregation with a small number of groups),
we expect this to remain true for scale factors $>128$.

\subsection{Compute Intensity}

\begin{table}
  \caption{Query complexity.}
  \label{tbl:computational-complexity}
  \vspace{-3ex}
  \centering
  \begin{tabular}{@{}lcS[table-format=3.1,table-auto-round]@{}}
    \toprule
    \textbf{Query}  & \textbf{Complexity} & \textbf{\#Ops/event}  \\
    \midrule
    \queryref{1}    & $1$                 & 1.0                   \\
    \queryref{2}    & $J$                 & 3.198597868458298     \\
    \queryref{3}    & $J$                 & 3.198597868458298     \\
    \queryref{4}    & $1 + J$             & 4.198597868458276     \\
    \queryref{5}    & $1 + \binom{M}{2}~~$  & 1.5871944717190243    \\
    \queryref{6}    & $~~1 + \binom{J}{3}$  & 42.79319653383031     \\
    \queryref{7}    & $(E + M) \cdot \sigma(J)$ & 1.516950186054388 \\
    \queryref{8}    & $E \cdot M + E + M + 1$   & 11.60497977798159 \\
    \bottomrule
  \end{tabular}
\end{table}

\begin{figure}
  \centering
  \includegraphics[scale=.7]{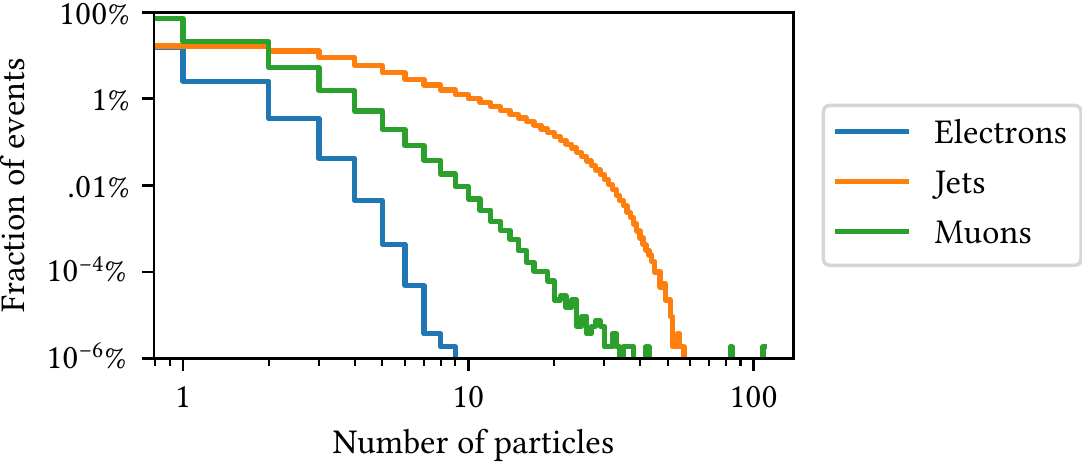}
  \caption{Distribution of number of particles per event.}
  \label{fig:particle-distribution}
\end{figure}

We now study the balance of compute and I/O in more detail.
We start with a complexity analysis of the queries.
Table~\ref{tbl:computational-complexity} gives a formula for each query
indicating how many records or record combinations
the query must explore for each event.
Note that this number is 1 for all scans not involving arrays
and hence for the traditional use cases with data in NF1.
$E$, $J$, and $M$ denote the number of electrons, muons, and jets, respectively;
$\sigma$ denotes the filter used in \queryref{7}.
While \queryref{1} does not access any particle array
and \queryref{2} to \queryref{4} only accesses one of them,
the remaining queries produce particle \emph{combinations}
as discussed in Section~\ref{sec:implementation:combinations}.
From the formulae, we can understand that the compute intensity of a query
depends on the distribution of the number of particles \emph{per event}.
Figure~\ref{fig:particle-distribution} shows this distribution
for the data set of the benchmark
and the three particle types used by the queries.
It shows that electrons generally occur in the low single-digit numbers,
muons generally occur more frequently and reach higher per-event occurrences,
and a significant fraction of the events consists of several dozen jets.
In these ``large'' events, exploring all combinations of three jets,
like required by \queryref{6}, may lead to a significant cost.
For example, a single event with 50 jets
has $\binom{50}{3} = \num{19600}$ combinations of three jets,
for each of which queries typically compute some distance
using a sequence of expensive geometric functions.
The last column of Table~\ref{tbl:computational-complexity} shows
how many records or record combinations the query must explore
on average for each event in the benchmark data set.
\queryref{6}, in particular, and to some degree \queryref{8}
are thus \emph{intrinsically} compute-intensive
and the running time of even highly tuned execution engines
are likely to be dominated by these computations rather than by I/O.

\begin{figure}
  \centering
  \includegraphics[scale=.7]{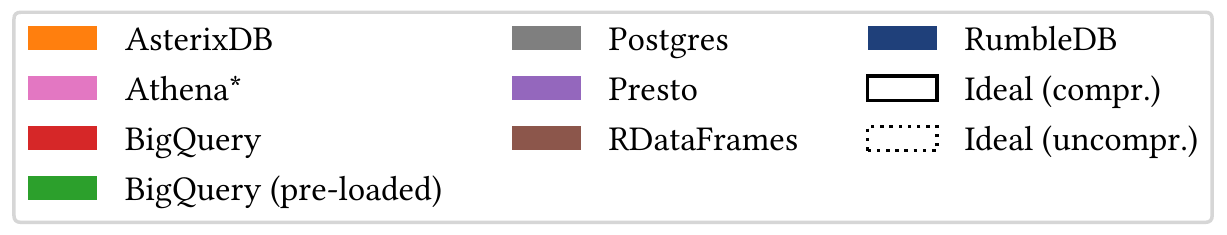}
  \begin{subfigure}[t]{\linewidth}
    \hfill\includegraphics[scale=.7]{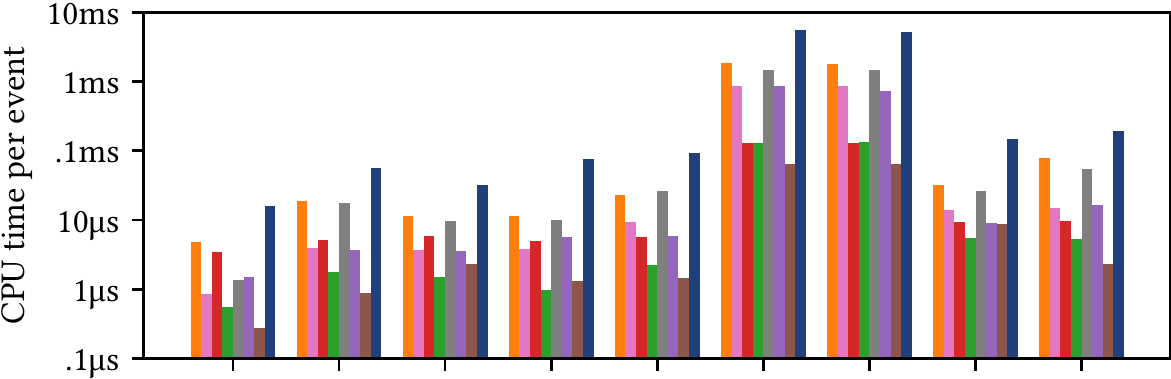}
    \vspace{-1ex}
    \caption{Average CPU time per event.}
    \label{fig:systems:cpu-time}
    \vspace{1ex}
  \end{subfigure}
  \begin{subfigure}[t]{\linewidth}
    \hfill\includegraphics[scale=.7]{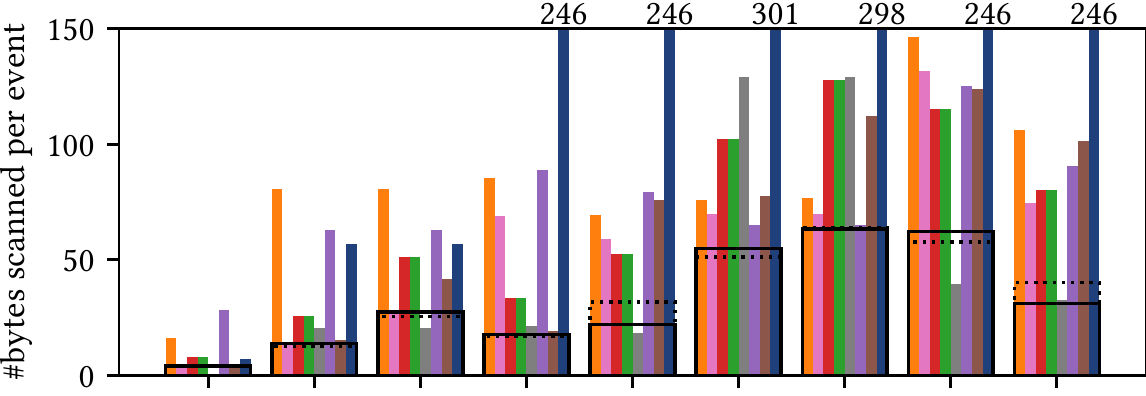}
    \vspace{-1ex}
    \caption{Average amount of data scanned per event.}
    \label{fig:systems:data-scanned}
    \vspace{1ex}
  \end{subfigure}
  \begin{subfigure}[t]{\linewidth}
    \hfill\includegraphics[scale=.7]{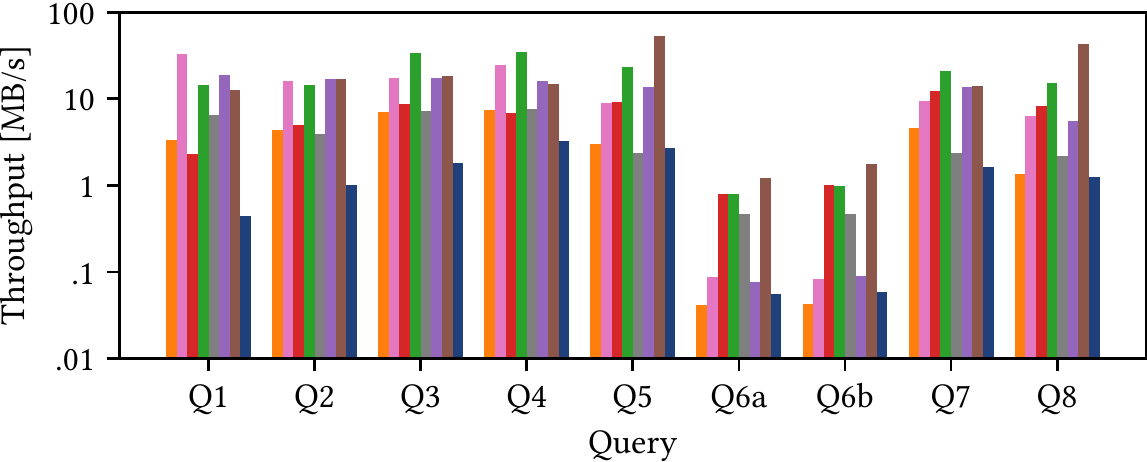}
    \vspace{-1ex}
    \caption{End-to-end processing throughput per logical core.}
    \label{fig:systems:scan-throughput}
  \end{subfigure}
  \caption{Analysis of compute/IO balance.}
  \label{fig:systems}
\end{figure}

To quantify the balance between I/O and compute further,
we compare three metrics across the queries and systems
in Figure~\ref{fig:systems}.
In Figure~\ref{fig:systems:cpu-time}, we show the average CPU time,
which we compute as the total number of seconds any logical core
spends on doing work for the query
and which excludes the wait time of these cores.
BigQuery, Presto, and RumbleDB report this metric in their statistics;
for the remaining self-hosted systems,
we use the CPU time as reported by the OS kernel.
As a best effort, we run the queries on Athena
against a version of the input that consists of a single row group
such that the query has to be processed sequentially
and then take the wall time as the CPU time;
this approximation may not be accurate and has to be taken with a grain of salt
(hence the * in the legend of the plot).
The plot shows that the CPU time follows similar trends
as the end-to-end running time reported above:
the ranking among the systems is the same
and \queryref{6}, \queryref{8}, \queryref{7}, and \queryref{5}
take the longest to compute.
This is also in line with the computational complexity of the queries
shown in Table~\ref{tbl:computational-complexity}.

In Figure~\ref{fig:systems:data-scanned}, we show the number of bytes
scanned \emph{per event}.
Most systems report that metric as well;
otherwise, we use statistics from the IO and networking subsystems of the OS.
We also show the ideal value for that metric,
once computed based on the column size as reported by the Parquet metadata,
once based on the number of column entries
times the size of the data type (\SI{4}{\byte} for most attributes).
We see that all systems scan significantly more data than they should:
As discussed before, BigQuery reports \SI{2}{\times} more
than it actually reads from storage due to its pricing model.
Also, AsterixDB, Athena, and Presto are not able
to push down projections into structs;
instead, they always read all fields
of any struct attribute accessed.
This is most likely due to a limitation
of the Java implementation of the Parquet format,
but it is not intrinsic to the format itself---%
the C++ implementation does not have this shortcoming.
RumbleDB does not seem to push \emph{any} projection into the scan
and thus reads the full file for all but the simplest queries.
RDataFrame also causes more bytes to be read than expected
even though the projections are manually specified by the user;
further work is needed to understand why.
We could not determine
why Postgres seems to read \emph{less} data than ideal in some queries;
it is possible that it caches part of the input in its buffer pool.

Finally, Figure~\ref{fig:systems:scan-throughput}
shows the scan throughput per core, computed as the amount of data scanned divided by the total CPU time.
This number reflects the balance of compute and I/O.
Comparing it with the typical I/O bandwidth obtainable from storage,
typically around \SIrange{50}{200}{\mega\byte\per\second} per core,
it indicates whether the queries are bound by I/O or by compute.
If we discount the numbers of BigQuery
by the \SI{2}{\times} inflation of its pricing model,
none of the systems comes close to the raw storage bandwidth:
BigQuery, Presto, and Athena achieve
only \SIrange{15}{30}{\mega\byte\per\second}
on \queryref{1} to \queryref{4};
BigQuery more or less maintains these numbers
also for \queryref{7} and \queryref{8},
while the others fall below \SI{10}{\mega\byte\per\second}.
On \queryref{6}, BigQuery and RDataFrames
achieve a mere \SI{1}{\mega\byte\per\second},
while AsterixDB, Athena, Presto, and RumbleDB
drop to as little as \SI{100}{\kilo\byte\per\second}.
This indicates that the systems are heavily compute-bound,
which is to some degree intrinsic to the queries
as shown by the complexity analysis above;
however, as the difference in performance among the systems indicates,
most systems are also significantly less efficient than possible,
often by one or even several orders of magnitude.

\vspace{-1ex}
\section{Related Work}
\label{sec:related-work}

Over the decades, many researchers have studied
the possibility of using database systems
for data-intensive scientific applications in general%
~\cite{10.1145/1084805.1084808}
and high-energy physics in particular%
~\cite{10.1145/304181.304229,2019disr.confE.223C,Pivarski2018,
       Limper2014,Karpathiotakis2014,Grossman1994,Fry1993,Dai2018,%
       Malon2011,Marstaller1993,Baden1991,Cranshaw2010,Shiers2011,%
       Kernert2015,Vassilev2015,Nowak2001,Malon1995,Binko1996,Bowen2000}.
The domain where these efforts have been most successful
is arguably that of astrophysics
with the Sloan Digital Sky Survey~\cite{Szalay2008}.
In most other areas, real-world adoption seems rather limited.

However, researchers in the HEP domain
do seem to see the need to look beyond their mainstream tools.
For instance, Pheasant~\cite{Amaral:2003be}
is a visual query language for expressing decay queries more easily.
A more recent effort aims to open up particle physics analysis tools
to the broader scientific community
via the Scikit-HEP project~\cite{Rodrigues_2019}.
Other works have looked into exploiting the efficiency of GPUs
for HEP analyses~\cite{Pompili_2016}.
Also the ROOT file format is subject to investigation:
There are also several studies~\cite{Pivarski2017,Blomer_2018}
of the performance of the ROOT data format that compare it
with general-purpose alternatives
such as Parquet, Avro, or Protocol Buffers,
as well as propositions for new file formats%
~\cite{Gutsche2020,Chang2018}.
Finally, several authors have proposed integrations%
~\cite{Laurelin2020,viktor_khristenko_2017_1034230,Meoni2018,%
       Sehrish2017,Baranowski2019}
with modern big-data systems such as Apache Spark.

There are several studies that compare query languages for certain domains.
For example, \textcite{Ong2014} analyze
eleven document-oriented query languages,
from which they eventually derive the design of SQL++,
the language that is part of the comparison of this paper.
While for a different use case,
the study follows a similar structure as ours.
A similar study compares high-level query languages
in the context of MapReduce~\cite{Stewart2011}.

% Here we present the final ideas and concluding remarks
\section{Conclusion}
\label{sec:conclusion}
We have evaluated
several general-purpose data processing systems
in terms of suitability of their query language,
absolute performance, scalability, and query price
in the context of High-energy Physics (HEP).
With the support for structured data types and arrays,
several SQL dialects can express HEP analyses reasonably well,
and languages for nested and heterogeneous data
allow for more natural query formulations.
However, the general-purpose data processing systems
are significantly less performant than the domain-specific ROOT framework---%
due to limited scalability and inefficient handling of
the data and queries relevant to HEP.

The observations of the study
suggest several avenues for future research:
On the one hand, efficiency on nested but \emph{homogeneous} data
needs to be improved in order to make database systems competitive---%
otherwise, the potential of their many advantages remains untapped.
On the other hand, to enable real-world adoption,
some further questions need to be solved:
how to expose the large body of physics libraries
offered by ROOT and similar frameworks in query languages,
how to integrate query languages with plotting and archival facilities,
and how to convince and prepare domain physicists
to adopt a new tool chain.

\begin{acks}
We thank Jim Pivarski for establishing the connection among the authors
and the various insightful discussions on HEP analyses,
as well as the respective developer teams of ROOT and AsterixDB
for their timely and thorough replies.
\end{acks}

\afterpage{\afterpage{\afterpage{\afterpage{\enlargethispage{-8.5cm}}}}}

\printbibliography

%\appendix
%\input{sections/appendix}

\end{document}